\documentclass[aps,nofootinbib,showpacs,preprintnumbers,amsmath,amssymb]{revtex4}
\def\bes{\begin{subequations}}
\def\ees{\end{subequations}}
\def\be{\begin{equation}}
\def\ee{\end{equation}}
\def\bea{\begin{eqnarray}}
\def\eea{\end{eqnarray}}
\def\ba{\begin{eqnarray}}
\def\ea{\end{eqnarray}}
\def\bear{\begin{array}}
\def\eear{\end{array}}
\def\p1sl{\displaystyle{\not}p_1}
\def\p2sl{\displaystyle{\not}p_2}
\def\pMsl{\displaystyle{\not}p_M}
\def\pMpsl{\displaystyle{\not}p_{M'}}

\newcommand{\K}{{\widetilde {\cal K}}}
\newcommand{\G}{{\overline {\Gamma}}}
\newcommand{\bG}{{\overline{\Gamma}}}

\usepackage{graphics}
\usepackage{graphicx}
\usepackage{dcolumn}
\usepackage{bm}
\usepackage{epsfig}
\usepackage{graphicx}
\usepackage{multirow}
\usepackage{dcolumn}
\usepackage{graphicx,epsfig}%

\begin{document}
\preprint{USM-TH-321}

\title{CP violation in lepton number violating semihadronic decays of
$K, D, D_s, B, B_c$}
\author{Gorazd Cveti\v{c}$^1$}
\email{gorazd.cvetic@usm.cl}
\author{C.~S.~Kim$^2$}
\email{cskim@yonsei.ac.kr}
\author{Jilberto Zamora-Sa\'a$^1$}
\email{jilberto.zamora@usm.cl}
\affiliation{$^1$
Department of Physics, Universidad T\'ecnica Federico Santa Mar\'ia, Casilla 110-V, Valpara\'iso, Chile\\
$^2$Department of Physics and IPAP, Yonsei University, Seoul 120-749, Korea}

\date{\today}

\begin{abstract}
We study the CP violation in lepton number violating
meson decays $M^{\pm} \to \ell_1^{\pm} \ell_2^{\pm} M^{' \mp}$,
where $M$ and $M^{'}$ are pseudoscalar mesons,
$M=K, D, D_s, B, B_c$ and $M^{'}=\pi, K, D, D_s$, and the charged
leptons are $\ell_1, \ell_2 = e, \mu$. It turns out that the CP-violating
difference $S_{-}(M) \equiv [\Gamma(M^- \to \ell_1^- \ell_2^- M^{' +})-\Gamma(M^+ \to \ell_1^+ \ell_2^+ M^{' -})]$ can become appreciable when 
two intermediate on-shell Majorana neutrinos $N_j$ ($j=1,2$)
participate in these decays. Our calculations show that the asymmetry becomes 
largest when the masses of $N_1$ and $N_2$ are almost degenerate, 
i.e., when the mass difference $\Delta M_N$ becomes comparable with the 
(small) decay widths $\Gamma_N$ of these neutrinos: $\Delta M_N \not\gg  \Gamma_N$.
We show that in such a case, the CP ratio ${\cal A}_{CP}(M) \equiv 
[\Gamma(M^- \to \ell_1^- \ell_2^- M^{' +})-\Gamma(M^+ \to \ell_1^+ \ell_2^+ M^{' -})]/[\Gamma(M^- \to \ell_1^- \ell_2^- M^{' +})+\Gamma(M^+ \to \ell_1^+ \ell_2^+ M^{' -})]$ becomes a quantity $\sim 1$.
The observation of CP violation
in these decays would be consistent with the existence of the well-motivated
$\nu$MSM model with two almost degenerate heavy neutrinos in the
mass range between $M_N \sim 0.1$-$10^1$ GeV.
\end{abstract}


\maketitle

\section{Introduction}
\label{intr}

At this moment, one of the main questions in neutrino physics is
unresolved: whether the neutrinos are Majorana or Dirac particles.
If the neutrinos are Dirac particles, 
the lepton number is conserved in all processes.
If the neutrinos are Majorana particles, i.e., if they
are indistinguishable from their antiparticles, 
the lepton number in the reactions involving them
can be violated. 
The main processes whose eventual detection would decide on the
nature of neutrinos are the neutrinoless double beta decays 
($0 \nu \beta \beta$) in nuclei \cite{0nubb}. Among other processes
which may reflect the character of neutrinos are specific scattering processes
\cite{scatt1,scatt2,scatt3,scatt4}, and rare meson decays
\cite{LittSh,DGKS,Ali,IvKo,GoJe,Atre,HKS,CDKK,CDK}.

Another important question is the value of the masses of neutrinos.
Neutrino oscillations were predicted a long time ago \cite{Pontecorvo}, 
under the assumption that neutrinos have masses. These oscillations
were later observed \cite{oscatm,oscsol,oscnuc}, leading to the
conclusion that the first three neutrinos have nonzero but very
light masses $\alt 1$ eV. They can be produced via a seesaw
mechanism \cite{seesaw}, where the light neutrinos have
masses $\sim {\cal M}_D^2/{\cal M}_R$ ($\alt 1$ eV),
where ${\cal M}_D$ is an electroweak scale or lower.
The heavy Majorana neutrinos in these seesaw scenarios are very heavy, 
with typical masses ${\cal M}_R \gg 1$ GeV,
and their mixing with active neutrino flavors is very
suppressed $\sim {\cal M}_D/{\cal M}_R$ ($\ll 1$). However,
scenarios exist \cite{scatt2,nuMSM,HeAAS,KS,AMP} 
where the heavy Majorana neutrinos can have
relatively low masses $\sim 1$ GeV and their mixings with active
neutrinos flavors can be larger than in the usual seesaw scenarios.

Another important question in neutrino physics is the strength (if any) of
the CP violation in the neutrino sector. It could be measured
by neutrino oscillations \cite{oscCP}. However, in this work we will 
investigate the possibility of detection of CP violation in the rare
lepton number violating (LNV) semihadronic decays of 
charged pseudoscalar mesons.

In general CP  violation
is expected in both cases of neutrinos being Dirac or Majorana particles. 
Nonetheless, in the Pontecorvo–-Maki–-Nakagawa–-Sakata (PMNS) 
mixing matrix \cite{Pontecorvo,MNS}
the number of possible CP-violating phases
is larger when the neutrinos are Majorana particles.
If $n$ is the number of neutrino generations, the number of CP-violating phases
is $n(n-1)/2$ in the Majorana case, and $(n-1)(n-2)/2$ in the Dirac case,
cf.~Ref.~\cite{Bilenky}.

In a recent work \cite{CKZ}, we investigated the
possibility of measuring the CP asymmetry in the rare
leptonic decays of charged pions $\pi^{\pm} \to e^{\pm} e^{\pm} \mu^{\mp} \nu$.
Both lepton number conserving (LNC) and lepton number violating (LNV)
processes contribute to these decays and to the CP violation.
We concluded that the CP violation is appreciable when 
these processes are mediated by two on-shell (Majorana or Dirac)
sterile neutrinos $N_1$ and $N_2$ 
(i.e., with masses between $106$ and $140$ MeV), and that the 
CP violation effect is largest when
these two neutrinos are almost degenerate in their masses.
It is interesting that such neutrinos fall within the regime
predicted by the $\nu$MSM model \cite{nuMSM,Shapo}. 
Further, they are not ruled out by experiments \cite{PDG2012,Atre}.

The $\nu$MSM model \cite{nuMSM,Shapo}
contains two almost degenerate sterile Majorana neutrinos with mass
between $100$ MeV and a few GeV, and in addition a light sterile Majorana 
neutrino of mass $\sim 10^1$ keV and the three very light neutrinos.
The model is well motivated because:
(a) it can explain simultaneously the pattern of light neutrino masses 
and oscillations; (b) it can explain the baryon asymmetry of the Universe;
(c) it provides a dark matter candidate. 
We refer to Refs.~\cite{nuMSMrev} 
for reviews, and to Refs.~\cite{CDSh} for the determination of the
allowed range of the sterile neutrinos of the $\nu$MSM model.
Remarkably, the tentative evidence of a dark matter line,
recently discussed in Refs.~\cite{DM}, falls into the regime
predicted for $\nu$MSM in Refs.~\cite{CDSh}.
It is interesting that the requirement that the lightest sterile neutrino be
the dark matter candidate reduces the parameters of the model
in such a way as to make the two heavier neutrinos nearly degenerate
in mass. This in turn, as demonstrated in Ref.~\cite{CKZ}, increases
significantly the possible effects of CP violation.

Moreover, the CERN-SPS has proposed a search of such
heavy neutrinos, Ref.~\cite{CERN-SPS}, 
in the leptonic and semihadronic decays of $D$, $D_s$ mesons. 
As argued in \cite{CERN-SPS} and in the works 
\cite{LittSh,DGKS,Ali,IvKo,GoJe,Atre,HKS,CDKK,CDK},
such rare decays can have appreciable rates to be detected
in future experiments (such as the experiment proposed at CERN-SPS).

In this work we investigate such rare semihadronic decays of
charged pseudoscalar mesons $M^{\pm} \to \ell_1^{\pm} \ell_2^{\pm} M^{' \mp}$,
where $M=K, D, D_s, B, B_c$ and $M^{'}=\pi, K, D, D_s$, and the charged
leptons are $\ell_1, \ell_2 = e, \mu$. These decays are
lepton number violating (LNV), hence the neutrinos mediating them
must be of Majorana type.
We focus on signals of CP violation in such processes, by
working in scenarios with two on-shell sterile neutrinos
$N_1$ and $N_2$, i.e., with masses $M_{N_j}$ in the intervals
$M_{M'}+ M_{\ell_2} < M_{N_j} < M_M - M_{\ell_1}$.
The signals of CP violation are represented by the
CP-violating difference
$S_{-}(M) \equiv [\Gamma(M^- \to \ell_1^- \ell_2^- M^{' +})-\Gamma(M^+ \to \ell_1^+ \ell_2^+ M^{' -})]$,
and alternatively by the usual CP ratio
 ${\cal A}_{CP}(M) \equiv 
[\Gamma(M^- \to \ell_1^- \ell_2^- M^{' +})-\Gamma(M^+ \to \ell_1^+ \ell_2^+ M^{' -})]/[\Gamma(M^- \to \ell_1^- \ell_2^- M^{' +})+\Gamma(M^+ \to \ell_1^+ \ell_2^+ M^{' -})]$.

In Sec.~\ref{sec:form} we describe the formalism for calculation
of the various decay widths.
The details of the calculation are given in Appendix \ref{app1};
and the details for the total decay widths $\Gamma_N(M_N)$ of the (heavy)
sterile Majorana neutrinos are given in Appendix \ref{app2}.
In Sec.~\ref{sec:ACPsum} we present the expressions for the
decay widths
$S_{\pm}(M) \equiv
[\Gamma(M^- \to \ell_1^- \ell_2^- M^{' +}) \pm \Gamma(M^+ \to \ell_1^- \ell_2^- M^{' +})]$
and for the mentioned CP ratio ${\cal A}_{CP}(M)$.
Additional details are given in Appendix \ref{app3}.
In Sec.~\ref{sec:acc} we discuss the acceptance factor due to the
(long) decay time of the on-shell sterile neutrinos, and the 
resulting effective (i.e., experimental) branching ratios
${\rm Br}^{\rm (eff)}(M)$ [$\propto S_{+}(M)$] and
${\cal A}_{\rm CP}(M){\rm Br}^{\rm (eff)}(M)$ [$\propto S_{-}(M)$], 
and present numerical results.
In Sec.~\ref{sec:concl} we summarize our results and present conclusions.

\section{The process and formalism for the LNV semihadronic decays of pseudoscalars}
\label{sec:form}

\begin{figure}[htb] 
\begin{minipage}[b]{.49\linewidth}
\includegraphics[width=70mm]{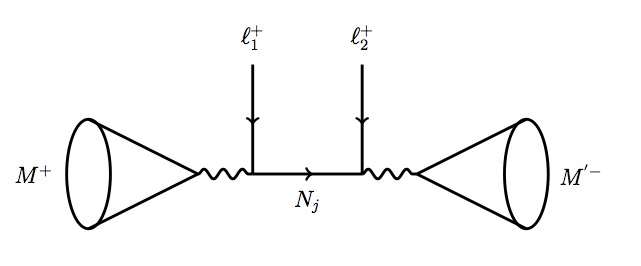}
\end{minipage}
\begin{minipage}[b]{.49\linewidth}
\includegraphics[width=70mm]{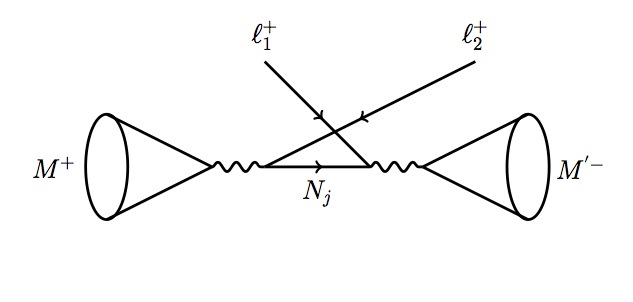}
\end{minipage}
\vspace{-0.4cm}
\caption{The lepton number violating decay $M^+ \to \ell_1^+ \ell_2^+ M^{' -}$, e.g. 
with $M=K$ and $M^{'}=\pi$: the direct (D) channel (the left-hand figure); the crossed (C) channel (the right-hand figure).}
\label{FigLV}
 \end{figure}

We consider the lepton number violating (LNV) processes, Fig.~\ref{FigLV},
$M^{\pm} \to \ell_1^{\pm} \ell_2^{\pm} M^{' \mp}$, where the two intermediate
Majorana neutrinos ($N_1$, $N_2$) are on-shell. 
The intermediate neutrinos have to be Majorana here because these processes 
violate lepton number.

In such a case, the topology of these tree level processes is 
like ``$s$-channel.'' The processes with (two-loop) ``$t$-channel'' topology 
are strongly suppressed \cite{IvKo}. The type of processes of Fig.~\ref{FigLV}, within the models with sterile neutrinos $N$ in the mass range of mesons, 
have been studied in several works, among them
Refs.~\cite{LittSh,DGKS,Ali,IvKo,GoJe,Atre,HKS,CDKK,CDK}.

We denote the mixing coefficient for the heavy mass eigenstate $N_j$
with the standard flavor neutrino $\nu_{\ell}$ ($\ell = e, \mu, \tau$) 
as $B_{\ell N_j}$ ($j=1,2)$.\footnote{
There exist also other notations for $B_{\ell N}$ in the literature, e.g.
$U_{\ell  N}$ in \cite{HKS}; $V_{\ell  N}$ in \cite{Atre}.} 
The relevant mixing relations in our notation are
\be
\nu_{\ell} = \sum_{k=1}^3 B_{\ell \nu_k} \nu_k +
\left( B_{\ell N_1} N_1 + B_{\ell N_2} N_2 \right) \ ,
\label{mix}
\ee
where $\nu_k$ ($k=1,2,3$) are the light mass eigenstates, and the
(unitary) PMNS matrix $B$ is in this scenario a $5 \times 5$ matrix.\footnote{
In our work, $B$ can be a $n \times n$ matrix with $n \geq 5$.
If $n > 5$, we implicitly assume that the additional sterile neutrinos 
($N_3$, etc.) have significantly less mixing than $N_1$ and $N_2$
with the active flavor (``light'') neutrino sector; one such framework
is $\nu$MSM \cite{nuMSM,Shapo,nuMSMrev}, with $n=6$.} 

We will use the phase conventions of the book Ref.~\cite{Bilenky}, i.e.,
all the CP-violating phases are incorporated in the PMNS matrix of mixing
elements.
The sum and difference of the decay widths,
$S_{\pm}(M) \equiv
[\Gamma(M^- \to \ell_1^- \ell_2^- M^{' +}) \pm \Gamma(M^+ \to \ell_1^- \ell_2^- M^{' +})]$,
of the processes of Fig.~\ref{FigLV}
will be appreciable only if the
two intermediate neutrinos $N_j$ are on-shell
\bea
(M_{M'} + M_{\ell_2}) &<& M_{N_j} < (M_{M}-M_{\ell_1}) \ , \ {\rm or/and}
\nonumber\\
(M_{M'} + M_{\ell_1}) &<& M_{N_j} < (M_{M}-M_{\ell_2}) \ .
\label{MNjint}
\eea

We will often use schematic notations for the decay widths of these rare processes:
\be
\Gamma(M^{\pm}) \equiv \Gamma(M^{\pm} \to \ell_1^{\pm} \ell_2^{\pm} M^{' \mp}) \ .
\label{not0}
\ee

These decay widths can be written in the form
\be
\Gamma(M^{\pm}) =  (2 - \delta_{\ell_1 \ell_2} ) \frac{1}{2!} \frac{1}{2 M_M} \frac{1}{(2 \pi)^5}
\int d_3 \; | {\cal T}(M^{\pm}) | ^2 \ ,
\label{GM1}
\ee
where $1/2!$ is the symmetry factor when the two charged leptons are equal.
Here, $|{\cal T}(M^{\pm})|^2$ is the absolute square 
(summed over the final helicities) of the sum of amplitudes 
from $N_1$ and $N_2$ neutrinos 
in the two channels $D$ (direct) and $C$ (crossed). We refer to Appendix \ref{app1}
for details.
In Eq.~(\ref{GM1}), $d_3$ denotes the integration over the three-particle final phase space
\be
d_3 \equiv \frac{d^3 {\vec p}_1}{2 E_{\ell_1}({\vec p}_1)} 
 \frac{d^3 {\vec p}_2}{2 E_{\ell_2}({\vec p}_2)}
 \frac{d^3 {\vec p}_{M'}}{2 E_{M'}({\vec p}_{M'})}
\delta^{(4)} \left( p_{M} - p_1 - p_2 - p_{M'} \right) \ .
\label{d3}
\ee
We denoted by $p_1$ and $p_2$ the momenta of $\ell_1$ and $\ell_2$
from the left and the right vertex of the direct channels, respectively 
(in the crossed channel $\ell_2$ couples to the left vertex), 
cf.~Fig.\ref{FigLV}.
The decay widths (\ref{GM1}) can then be written as a double sum
over the contributions of $N_i$ and $N_j$ exchanges ($i,j=1,2$), with the mixing effects
factored out
\ba
\Gamma(M^{\pm}) &=&  (2 - \delta_{\ell_1 \ell_2} )
\sum_{i=1}^2 \sum_{j=1}^2 k_i^{(\pm)} k_j^{(\pm)*} 
{\big [}
\bG(DD^{*})_{ij} + \bG(CC^{*})_{ij}
+ \bG_{\pm}(DC^{*})_{ij} + \bG_{\pm}(CD^{*})_{ij} {\big ]} \ ,
\label{GM2}
\ea
where $k_j^{(\pm)}$ are the corresponding mixing factors
\be
k_j^{(-)} = B_{\ell_1 N_j} B_{\ell_2 N_j} \ ,\qquad  k_j^{(+)}= (k_j^{(-)})^{*} \ ,
\label{kj}
\ee
and $\bG_{\pm}(XY^{*})_{ij}$ are the normalized (i.e., without the mixing) contributions 
of $N_i$ exchange in the $X$ channel and complex-conjugate of the $N_j$ exchange
in the $Y$ channel ($X,Y=C, D$) 
\be
\bG_{\pm}(XY^{*})_{ij} \equiv K^2 \; \frac{1}{2!} \frac{1}{2 M_M} \frac{1}{(2 \pi)^5}
\int d_3 \; P_i(X) P_j(Y)^{*} M_{N_i} M_{N_j} T_{\pm}(X) T_{\pm}(Y)^{*} \ .
\label{bGXY}
\ee
Here, $T_{\pm}(X)$ ($X=D,C$) are the relevant parts of the amplitude in the $X$ channel
which appear also in the total decay amplitudes ${\cal T}_{\pm}$ 
(see Appendix \ref{app1}),\footnote{
Since $|T_{+}(D)|^2 = |T_{-}(D)|^2$ and  $|T_{+}(C)|^2 = |T_{-}(C)|^2$, we
omitted the subscripts $\pm$ from the $DD^{*}$ and $CC^{*}$ contribution
terms $\bG(DD^{*})_{ij}$ and $\bG(CC^{*})_{ij}$ in Eq.~(\ref{GM2}).}
and $P_j(X)$ ($X=D,C$) are the propagators of the intermediate neutrinos 
$N_j$ in the two channels
\bes
\label{Pj}
\ba
P_j(D) &=& \frac{1}{\left[ (p_{M}-p_1)^2 - M_{N_j}^2 + i \Gamma_{N_j} M_{N_j} \right]} \ ,
\label{PjD}
\\
P_j(C) &=& \frac{1}{\left[ (p_{M}-p_2)^2 - M_{N_j}^2 + i \Gamma_{N_j} M_{N_j} \right]} \ .
\label{PjC}
\ea
\ees
The overall constant $K^2$ appearing in Eqs.~(\ref{bGXY}) is
\be
K^2 = G_F^4 f_{M}^2 f_{M'}^2  |V_{Q_u Q_d} V_{q_u q_d}|^2 \ ,
\label{Ksqr}
\ee
where $f_{M}$ and $f_{M'}$ are the decay constants of $M^{\pm}$ and $M^{' \mp}$,
and $V_{Q_u Q_d}$ and $V_{q_u q_d}$ are the CKM elements corresponding to $M^{\pm}$ and $M^{' \mp}$
(the valence quark content of $M^+$ is $Q_u {\bar Q}_d$; of $M^{' +}$ is $q_u {\bar q}_d$).

Several symmetry relations exist among the normalized decay widths
$\bG_{\pm}(XY^{*})_{ij}$, as given in Eqs.~(\ref{symm})-(\ref{symmadd}) in Appendix
\ref{app1}. The most important symmetry property is 
that the $(2 \times 2)$ matrices
$\bG(DD^{*})$ and $\bG(CC^{*})$ are self-adjoint (and even equal if $\ell_1=\ell_2$).
The matrices $\bG_{\pm}(DC^{*})$ and $\bG_{\pm}(CD^{*})$, which represent the
(normalized) $D$-$C$ channel interference contributions to
the decay widths $\Gamma(M^{\pm})$, will turn out to be several orders
of magnitude smaller than the $\bG(DD^{*})$ and $\bG(CC^{*})$ matrices.

In our calculations we will also need to know the total decay width 
$\Gamma(N_j \to {\rm all}) \equiv \Gamma_{N_j}$ of the two Majorana
neutrinos $N_j$ as a function of the mass $M_{N_j}$, or more specifically, 
the corresponding mixing factor $\K_j$. The width $\Gamma_{N_j}$ can be
written as
\begin{equation}
\Gamma_{N_j} = \K_j \G(M_{N_j}) \ ,
\label{GNwidth}
\end{equation}
where
\begin{equation}
 \G(M_{N_j}) \equiv \frac{G_F^2 M_{N_j}^5}{96 \pi^3} \ ,
\label{barG}
\ee
and the factor $\K_j$ includes the heavy-light mixing factors dependence
\begin{equation}
\K_j(M_{N_j}) \equiv \K_j = {\cal N}_{e N_j} \; |B_{e N_j}|^2 + {\cal N}_{\mu N_j} \; |B_{\mu N_j}|^2 + {\cal N}_{\tau N_j} \; |B_{\tau N_j}|^2 ) \ ,
\quad (j=1,2) \ .
\label{calK}
\end{equation}
Here, ${\cal N}_{\ell N}(M_N) \equiv {\cal N}_{\ell N}$ ($\ell = e, \mu, \tau$) are
the effective mixing coefficients; they are numbers $\sim 10^0$-$10^1$
which depend on the mass $M_N$ of the Majorana neutrino $N$ ($N=N_1, N_2$).  
In Appendix \ref{app2} we write down the relevant formulas for the calculation of these
coefficients. The results of these calculations are given in 
Fig.~\ref{FigcNellN}, for the here relevant neutrino mass interval
$0.1 \ {\rm GeV} < M_N < 6.3 \ {\rm GeV}$. Some additional remarks
are given in Appendix \ref{app2}.
\begin{figure}[htb]
\centering\includegraphics[width=100mm]{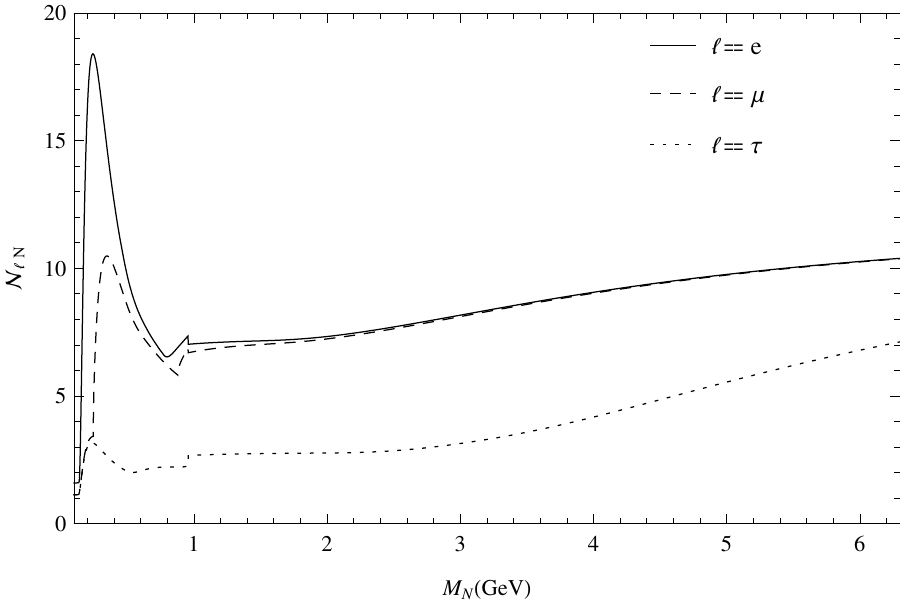}
\vspace{-0.4cm}
\caption{The effective mixing coefficients ${\cal N}_{\ell N}$ ($\ell = e, \mu, \tau$)
appearing in Eqs.~(\ref{GNwidth})-(\ref{calK}), as a function of the mass $M_N$ of
the Majorana neutrino $N$. See the text and Appendix \ref{app2} for details.}
\label{FigcNellN}
\end{figure}

On the other hand, the present upper bounds for 
the squares $|B_{\ell N}|^2$ of the heavy-light mixing matrix 
elements, in our range of interest 
$0.1 \ {\rm GeV} < M_N < 6.3 \ {\rm GeV}$, 
can be inferred from Ref.~\cite{Atre} 
(and references therein). 
The present upper bounds for $|B_{e N}|^2$, in the mentioned
range of $M_N$, are largely determined by the 
neutrinoless double beta decay experiments
\cite{Benes,Belanger} ($0\nu\beta\beta$). The upper bounds for
$|B_{\mu N}|^2$ come from searches of peaks in the spectrum of $\mu$ in
pion and kaon decays \cite{Kusenko} 
and from decay searches
\cite{Kusenko,beamdump,l3etal,delphi}. 
The upper bounds for $|B_{\tau N}|^2$ come from 
CC interactions (if $\tau$ is produced) and from NC interactions
\cite{delphi,noch}. In Table \ref{T1} we present the upper bounds on
$|B_{\ell N}|^2$ for specific chosen values of $M_N$ in the mentioned 
integral. 
\begin{table}
\caption{Present upper bounds for the squares $|B_{\ell N}|^2$ of the
heavy-light mixing matrix elements, for various specific values of $M_N$.}
\label{T1}
\begin{tabular}{| c | c | c | c |}
\hline
\bf{$M_N [GeV]$} & $|B_{eN}|^2$ & $|B_{\mu N}|^2$ & $|B_{\tau N}|^2$ \\
\hline
0.1 & $(1.5\pm 0.5)\times10^{-8} $ & $(6.0\pm 0.5)\times10^{-6}$ & $(8.0\pm 0.5)\times10^{-4}$ \\
\hline
0.3 & $(2.5\pm 0.5)\times10^{-9}$ & $(3.0\pm 0.5)\times10^{-9}$ & $(1.5\pm 0.5)\times10^{-1}$ \\
\hline
0.5 & $(2.0\pm 0.5)\times10^{-8}$ & $(6.5\pm 0.5)\times10^{-7}$ & $(2.5\pm 0.5)\times10^{-2}$ \\
\hline
0.7 & $(3.5\pm 0.5)\times10^{-8}$ & $(2.5\pm 0.5)\times10^{-7}$ & $(9.0\pm 0.5)\times10^{-3}$ \\
\hline
1.0 & $(4.5\pm 0.5)\times10^{-8}$ & $(1.5\pm 0.5)\times10^{-7}$ & $(3.0\pm 0.5)\times10^{-3}$ \\
\hline
2.0 & $(1.0\pm 0.5)\times10^{-7}$ & $(2.5\pm 0.5)\times10^{-5}$ & $(3.0\pm 0.5)\times10^{-4}$ \\
\hline
3.0 & $(1.5\pm 0.5)\times10^{-7}$ & $(2.5\pm 0.5)\times10^{-5}$ & $(4.5\pm 0.5)\times10^{-5}$ \\
\hline
4.0 & $(2.5\pm 0.5)\times10^{-7}$ & $(1.5\pm 0.5)\times10^{-5}$ & $(1.5\pm 0.5)\times10^{-5}$ \\
\hline
5.0 & $(3.0\pm 0.5)\times10^{-7}$ & $(1.5\pm 0.5)\times10^{-5}$ & $(1.5\pm 0.5)\times10^{-5}$ \\
\hline
6.0 & $(3.5\pm 0.5)\times10^{-7}$ & $(1.5\pm 0.5)\times10^{-5}$ & $(1.5\pm 0.5)\times10^{-5}$ \\
\hline
\end{tabular}
\end{table}
The upper bounds have in some cases strong dependence on the
precise values of $M_N$, and for further details we refer to the
corresponding figures in Ref.~\cite{Atre}.

\section{The decay widths and CP asymmetry for the LNV semihadronic decays of pseudoscalars}
\label{sec:ACPsum}

Here we will use the results of Sec.~\ref{sec:form}, and a combination
of analytic and numerical evaluations, in order to obtain the results for
the decay widths $S_{\pm}$
and the CP asymmetry ratios ${\cal A}_{\rm CP}$ 
of the discussed semihadronic LNV decays of pseudoscalar mesons $M^{\pm}$ 
\ba
S_{\pm}(M) & \equiv &  \Gamma(M^-) \pm \Gamma(M^+) \ ,
\label{Brdef}
\\
{\cal A}_{\rm CP}(M) & \equiv & \frac{S_{-}(M)}{S_{+}(M)}
 \equiv \frac{ \Gamma(M^-) - \Gamma(M^+)}{ \Gamma(M^-) + \Gamma(M^+)} \ ,
\label{Adef}
\ea
where we use the notations of 
Eq.~(\ref{not0}). $S_{+}(M)$ represents the total (sum) of the decay widths
of $M^+$ and $M^-$ for these rare LNV decays, while
$S_{-}(M)$ is the corresponding (CP-violating) difference. 
The ratio ${\cal A}_{\rm CP}(M)$
in Eq.~(\ref{Adef}) is the usual measure of the
relative CP violation effect.
We adopt the convention $M_{N_2} > M_{N_1}$,
and introduce the following notations related
with the heavy-light neutrino mixing elements $B_{\ell_1 N_j}$ and $B_{\ell_2 N_j}$
and their phases:
\bes
\label{not}
\ba
\kappa_{\ell_1} & = & \frac{|B_{\ell_1 N_2}|}{|B_{\ell_1 N_1}|} \ ,
\quad
\kappa_{\ell_2} =   \frac{|B_{\ell_2 N_2}|}{|B_{\ell_2 N_1}|} \ ,
\label{kap}
\\
B_{\ell_k N_j} & \equiv & |B_{\ell_k N_j}| e^{i \phi_{kj}}  \qquad (k,j = 1,2) \ ,
\label{phi}
\\
\theta_{ij} & = & (\phi_{1i} + \phi_{2i} - \phi_{1j} - \phi_{2j}) \quad
(i,j=1,2) \ .
\label{theta}
\ea
\ees
For example, if $\ell_1=\ell_2=\mu$, then 
$\theta_{21}=2 (\phi_{\mu 2} - \phi_{\mu 1})= 2 ( {\rm arg}(B_{\mu N_2})
- {\rm arg}(B_{\mu N_1}))$.
Here we will not write explicitly the $D$-$C$ channel interference contributions
to the quantities (\ref{Brdef})-(\ref{Adef}), as our numerical calculations
give us for them contributions which are several orders of magnitude
smaller that the contributions from the $D$ channel and from the $C$ channel.

The resulting sums  $S_{+}(M) \equiv  \left( \Gamma^(M^-) + \Gamma(M^+) \right)$ 
of the decay widths can then be written
in terms of only the normalized decay widths $\bG(XX^{*})_{11}$, $\bG(XX^{*})_{22}$ 
and ${\rm Re} \bG(XX^{*})_{12}$ (where $X=D; C$), 
and in terms of the phase difference $\theta_{21}$
\bea
S_{+}(M) & \equiv & \left( \Gamma(M^-) + \Gamma(M^+) \right)
\nonumber\\
&= &
2 (2 - \delta_{\ell_1 \ell_2} ) |B_{\ell_1 N_1}|^2 |B_{\ell_2 N_1}|^2
{\bigg \{} 
\bG(DD^{*})_{11}
\left[ 1 + \kappa_{\ell_1}^2 \kappa_{\ell_2}^2 \frac{\bG(DD^{*})_{22}}{\bG(DD^{*})_{11}}
+ 2 \kappa_{\ell_1} \kappa_{\ell_2} \cos \theta_{21} \delta_1 \right]
\nonumber\\
&& 
+ \bG(CC^{*})_{11}
\left[ 1 + \kappa_{\ell_1}^2 \kappa_{\ell_2}^2 \frac{\bG(CC^{*})_{22}}{\bG(CC^{*})_{11}}
+ 2 \kappa_{\ell_1} \kappa_{\ell_2} \cos \theta_{21} \delta_1 \right]
+ (D-C \; {\rm terms})
{\bigg \}} 
\label{Spl} \ ,
\eea
where we used the notations (\ref{not}), and the quantity $\delta_1$
measures the effect of $N_1$-$N_2$ overlap contributions
\be
\delta_j \equiv \frac{{\rm Re} \bG(XX^{*})_{12}}{ \bG(XX^{*})_{jj}} \ , \quad 
(X=D; C; \quad j=1; 2) \ .
\label{delX}
\ee
It is expected that $\delta_j \approx 0$ when
$\Delta M_N \gg \Gamma_{N_j}$ because in such a case the overlap (interference)
effects of the $N_1$ and $N_2$ exchanges are expected to be absent due to
a large distance between the two ``bumps'' of the neutrino propagators.
Numerical evaluations confirm this expectation and confirm that
$\delta_j$ is practically independent of the channel $X=D, C$
(see later on in this Section).

The (CP-violating) difference  
$S_{-}(M) \equiv  \left( \Gamma(M^-) - \Gamma(M^+) \right)$ of the LNV
rare decays is  
\bea
\lefteqn{
S_{-}(M) \equiv \left( \Gamma(M^-) - \Gamma(M^+) \right)
}
\nonumber\\
&= &
4 (2 - \delta_{\ell_1 \ell_2} ) |B_{\ell_1 N_1}| |B_{\ell_2 N_1}| |B_{\ell_1 N_2}| |B_{\ell_2 N_2}|
\left\{ \sin \theta_{21} \left[
{\rm Im} \bG(DD^{*})_{12} + {\rm Im} \bG(CC^{*})_{12} \right]
+  (D-C \; {\rm terms})
\right\} \ .
\label{Smi}
\eea
We can see that CP violation in these decays is proportional to
the CP-odd phase difference $\theta_{21}$ defined in Eq.~(\ref{theta}). The
other factor in this CP violation is the imaginary part of
$\bG(DD^{*})_{12}+\bG(CC^{*})_{12}$; this factor will be
investigated later on in this Section.

The decay widths $\Gamma_{N_j}$ are very small in comparison with the
masses $M_{N_j}$, due to the mixing suppression,
cf.~Eqs.~(\ref{GNwidth}-\ref{calK}) (in general $\Gamma_{N_j} \ll 1$ eV).
Therefore, the 
absolute value of the square of the intermediate neutrino propagator
can be approximated to a high degree of accuracy by the delta
function
\ba
|P_j(D)|^2 &=&
\left | \frac{1}{(p_{M}-p_1)^2-M^{2}_{N_j}+i \Gamma_{N_j} M_{N_j}} \right | ^2
\nonumber\\
&\approx &
\frac{\pi}{M_{N_j} \Gamma_{N_j}} \delta((p_{M}-p_1)^2-M^{2}_{N_j})\ ;
\quad ( j=1,2; \; \Gamma_{N_j} \ll M_{N_j} ) \ ,
\label{P1P1}
\ea
and analogous equation for $|P_j(C)|^2$.
Therefore, in the integration $d_3$, the part of integration $d p_N^2$
($p_N=p_{M}-p_1$ in $D$ channel; $p_N=p_{M}-p_2$ in $C$ channel) becomes
a trivial integration over a delta function, and the expressions for
the diagonal elements $\bG(DD^{*})_{jj}$ and $\bG(CC^{*})_{jj}$ can be
calculated analytically, cf.~Appendix \ref{app3}
 \ba
\bG(DD^{*})_{jj} & = &
 \frac{K^2 M_M^5}{128 \pi^2} \frac{M_{N_j}}{\Gamma_{N_j} } 
\lambda^{1/2}(1, x_j,x_{\ell_1}) 
\lambda^{1/2} \left( 1, \frac{x^{'}}{x_j},\frac{x_{\ell_2}}{x_j} \right)
Q(x_j; x_{\ell_1}, x_{\ell_2},x^{'}) \quad (j=1 \; {\rm or} \; j=2) \ ,
\label{GDD}
\ea
and $\bG(CC^{*})_{jj}$ is obtained from the expression (\ref{GDD}) by the
simple exchange $x_{\ell_1} \leftrightarrow x_{\ell_2}$
\ba
\bG(CC^{*})_{jj} & = & \bG(DD^{*})_{jj}(x_{\ell_1} \leftrightarrow x_{\ell_2}) \ .
\label{GCC}
\ea
In Eq.~(\ref{GDD})
we used the notations
\bes
\label{notGDD}
\ba
\lambda(y_1,y_2,y_3) & = & y_1^2 + y_2^2 + y_3^2 - 2 y_1 y_2 - 2 y_2 y_3 - 2 y_3 y_1 \ ,
\label{lambda}
\\
x_j &=& \frac{M_{N_j}^2}{M_M^2} \ , \quad
x_{\ell_s} =  \frac{M_{\ell_s}^2}{M_M^2} \ , \quad
x^{'} =\frac{M_{M'}^2}{M_M^2} \ ,  \quad (j=1,2; \; \ell_s=\ell_1, \ell_2) \ ,
\label{xjs}
\ea
\ees
and the function $Q(x_j; x_{\ell_1}, x_{\ell_2},x^{'})$ is given in Appendix \ref{app3}.
In the special case $\ell_1=\ell_2$, the expression for $\bG(DD^{*})_{jj}$ is
somewhat simpler and can be deduced, e.g., from Ref.~\cite{CDKK}. 
The expressions (\ref{GDD}) and (\ref{GCC}) are used in the evaluation of
the sum $S_{+}(M)$,  Eq.~(\ref{Spl}), of the rare decay widths of $M^{\pm}$.
In Eq.~(\ref{Spl}), the contributions of the $N_1$-$N_2$ overlap
effects are parametrized in the function $\delta_1$ defined in
Eq.~(\ref{delX}), and will be evaluated later on numerically.

In order to evaluate the CP-violating difference $S_{-}(M)$, Eq.~(\ref{Smi}), 
of the rare decay widths $M^{\pm}$, 
the evaluation of the quantity ${\rm Im} \bG(XX^{*})_{12}$
($X=D; C$) is of central importance. In the integrand of ${\rm Im} \bG(XX^{*})_{12}$
we have, according to Eq.~(\ref{bGXY}), as factor the following
combination of the propagators of $N_1$ and $N_2$:
\bes
\label{ImP1P2gen}
\ba
{\rm Im} P_1(D) P_2(D)^{*}
&= &
\frac{
\left( p_N^2 - M_{N_1}^2 \right)  \Gamma_{N_2} M_{N_2}
- \Gamma_{N_1} M_{N_1} \left( p_N^2 - M_{N_2}^2 \right)
}
{
\left[ \left( p_N^2 - M_{N_1}^2 \right)^2 + \Gamma_{N_1}^2 M_{N_1}^2
\right]
\left[ \left( p_N^2 - M_{N_2}^2 \right)^2 + \Gamma_{N_2}^2 M_{N_2}^2
\right]
}
\label{ImP1P2ex}
\\
& \approx &
\mathcal{P} \left ( \frac{1}{p^{2}_{N}-M^{2}_{N_1}} \right )
\pi\ \delta (p^{2}_{N}-M^{2}_{N_2})
-
\pi\ \delta (p^{2}_{N}-M^{2}_{N_1})
\mathcal{P} \left ( \frac{1}{p^{2}_{N}-M^{2}_{N_2}} \right )
\label{ImP1P2a}
\\
& = & \frac{\pi}{M^{2}_{N_2}-M^{2}_{N_1}} \left [ 
\delta  ( p^{2}_{N}-M^{2}_{N_2})+ \delta  ( p^{2}_{N}-M^{2}_{N_1})  \right ] \ ,
\label{ImP1P2b}
\ea
\ees
where we have $p_N=(p_M-p_1)$ in the direct (D) channel. 
In Eqs.~(\ref{ImP1P2a})-(\ref{ImP1P2b}) we assumed 
$\Gamma_{N_j} \ll | \Delta M_N | \equiv M_{N_2}-M_{N_1}$. The expression
(\ref{ImP1P2gen}) has formally the same structure with Dirac delta functions
as Eq.~(\ref{P1P1}), but the factors in front of these Dirac delta
functions are different now. Hence we can perform the integration over the
final particle  phase space in the same way, but now under the
more stringent assumption $\Gamma_{N_j} \ll | \Delta M_N |$ (and not just:
$\Gamma_{N_j} \ll M_{N_j}$ which is always fulfilled),\footnote{
We note that this mechanism is central to the CP violation effects
in the considered LNV semihadronic decays of charged pseudoscalar mesons.
This mechanism was presented in Ref.~\cite{CKZ} and applied there
to the CP violation of the rare leptonic decays of charged pions.}
leading to the result
\bes
\label{ImG12}
\ba
{\rm Im} \bG(DD^{*})_{12} &=&
\eta \; 
\frac{K^2 M_M^5}{128 \pi^2} 
\frac{M_{N_1} M_{N_2}}{(M_{N_2}+M_{N_1}) \Delta M_N}
\sum_{j=1}^2
\lambda^{1/2}(1, x_j,x_{\ell_1}) \ ,
\lambda^{1/2} \left( 1, \frac{x^{'}}{x_j},\frac{x_{\ell_2}}{x_j} \right)
Q(x_j; x_{\ell_1}, x_{\ell_2},x^{'}) \ ,
\label{ImGDD12}
\\
{\rm Im} \bG(CC^{*})_{12} & = & 
{\rm Im} \bG(DD^{*})_{12}(x_{\ell_1} \leftrightarrow x_{\ell_2}) \ ,
\label{ImGCC12}
\ea
\ees
where we denoted $\Delta M_N \equiv M_{N_2} - M_{N_1} > 0$.
In Eqs.~(\ref{ImG12}) we introduced an overall factor $\eta$ which accounts for the
effects $\Delta M_N \not\gg  \Gamma_N$, i.e., for the situation when
the approximation (\ref{ImP1P2a}) of ${\rm Im} P_1(D) P_2(D)^{*}$ in terms
of Dirac delta functions in not justified. Later on in this Section, we will evaluate
numerically the factor $\eta$. When $\Delta M_N \gg \Gamma_{N_j}$,
i.e., when the identity (\ref{ImP1P2a}) can be applied, the factor $\eta$
is equal to unity, $\eta=1$.

The normalized decay matrix elements $\bG(XY^{*})_{ij}$, Eq.~(\ref{bGXY}),
were evaluated also numerically, by versions of Monte Carlo integration,
independently by the two authors, using finite (small) widths
$\Gamma_{N_j}$ in the propagators. We confirmed numerically the
analytic expression (\ref{GDD}) for
$\bG^{(X)}(DD^{*})_{jj}$ ($ \propto 1/\Gamma_{N_j}$),
and the analytic expression (\ref{ImG12}) with
$\eta=1$ for ${\rm Im} \bG(DD^{*})_{12}$ ($\propto 1/\Delta M_N$)
when $\Delta M_N \gg \Gamma_{N_j}$.

Further, our numerical evaluations lead us to the conclusion that
the direct-crossed channel ($DC^{*}$ and $CD^{*}$) interference contributions to
the sum and the difference of the rare decay widths $S_{\pm}(M)$ of $M^{\pm}$
are by several orders of magnitude smaller that the
corresponding direct ($DD^{*}$) and crossed ($CC^{*}$) channel contributions
to these quantities, in all cases.\footnote{
For example, when $M^{\pm}=K^{\pm}$ and $M^{' \mp}=\pi^{\mp}$, 
and we choose in numerical calculation
$\Gamma_N \sim 10^{-3} \ {\rm GeV} \sim \Delta M_N$, the 
 $\bG(DD^{*})_{ij}$ and $\bG(CC^{*})_{ij}$ contributions are by about
two orders of magnitude larger than the
$D$-$C$ interference contributions $\bG_{\pm}(DC^{*})_{ij}$.
When $\Gamma_N$ and $\Delta M_N$ are decreased further
($\Gamma_N \sim \Delta M_N$), the
 $\bG(DD^{*})_{ij}$ and $\bG(CC^{*})_{ij}$ contributions increase
(they are $\propto 1/\Gamma_N$, or $\propto 1/\Delta M_N$), while
the $D$-$C$ interference contributions $\bG_{\pm}(DC^{*})_{ij}$ remain
approximately unchanged and become thus relatively insignificant.} 

In addition, our numerical evaluations give us values of the
parameters $\delta_j$  of Eq.~(\ref{delX}), and of the
$\eta$ correction parameters of Eqs.~(\ref{ImG12}). In the cases
when $\Delta M_N \not\gg \Gamma_{N_j}$, these values differ appreciably from their
limiting values $\delta_j=0$ and $\eta=1$ of the $\Delta M_N \gg \Gamma_{N_j}$ limit.
It turns out that the parameters $\delta_j$ are 
practically independent of the channel contribution considered
($DD^{*}$ or $CC^{*}$) and of the type of pseudoscalar mesons
($M^{\pm}$, $M^{' \mp}$) and of the light leptons
($\ell_1, \ell_2 = e, \mu$) involved in the considered decays, and the same is
true for the parameter $\eta$. Further, numerical calculations
show that, 
in the considered case $\Delta M_N \not\gg \Gamma_{N_j}$ (i.e., when $N_1$ and $N_2$
are almost degenerate), the parameters $\eta$ and $\delta \equiv 
(1/2) (\delta_1 + \delta_2)$ are functions of only one parameter
$y \equiv \Delta M_N/\Gamma_N$,
where $\Delta M_N \equiv M_{N_2} - M_{N_1}$ ($> 0$)
and $\Gamma_N = (1/2)(\Gamma_{N_1} +\Gamma_{N_2})$
\bes
\label{etadel}
\ba
\eta &=& \eta(y) \ , \quad y \equiv \frac{\Delta M_N}{\Gamma_N} \ ,
\quad \Gamma_N \equiv \frac{1}{2} (\Gamma_{N_1} + \Gamma_{N_2}) \ ,
\label{etadel1}
\\
\delta &=&\delta(y) \ , \quad \delta \equiv  \frac{1}{2} (\delta_1 + \delta_2) \ ,
\quad
\frac{\delta_1}{\delta_2} = \frac{\bG(DD^*)_{22}}{\bG(DD^*)_{11}} =
\frac{\Gamma_{N_1}}{\Gamma_{N_2}} =
\frac{\K_1}{\K_2} \ .
\label{etadel2}
\ea
\ees
The numerical integration gives us these values, which are tabulated in
Table \ref{T2} as a function of $y$. 
\begin{table}
\caption{Values of $\delta(y)$, $\eta(y)$, and $\eta(y)/y$
correction parameters as a function of $y \equiv  \Delta M_N/\Gamma_N$.}
\label{T2}
\begin{tabular}{ll|lll}
$y \equiv \frac{\Delta M_N}{\Gamma_N}$ & $\log_{10} y$ &
$\delta(y)$ & $\eta(y)$ & $\frac{\eta(y)}{y}$
\\
\hline
1.00 & 0.000 & $0.500 \pm 0.004$ & $0.500 \pm 0.001$ & $0.500 \pm 0.001$
\\
1.25 & 0.097 & $0.390 \pm 0.003$ & $0.610 \pm 0.003$ & $0.488 \pm 0.002$
\\
1.67 & 0.222 & $0.264 \pm 0.003$ & $0.736 \pm 0.002$ & $0.441 \pm 0.001$
\\
2.50 & 0.398 & $0.138 \pm 0.001$ & $0.862 \pm 0.001$ & $0.345 \pm  0.001$
\\
5.00 & 0.699 & $0.038 \pm 0.001$ &$ 0.962 \pm 0.002$ & $0.192 \pm 0.001$
\\
10.0 & 1.000 & $0.0098 \pm 0.0010$ & $0.990 \pm 0.002$ & $0.0990 \pm 2 \times 10^{-4}$
\end{tabular}
\end{table}
The uncertainties indicate the
numerical uncertainties and the small variations from the various
considered LNV semihadronic decays 
$M^{\pm} \to \ell_1^{\pm} \ell_2^{\pm} M^{' \mp}$,
where $M$ and $M^{'}$ are pseudoscalar mesons,
$M=K, D, D_s, B, B_c$ and $M^{'}=\pi, K, D, D_s$, and the charged
leptons are $\ell_1, \ell_2 = e, \mu$. It is interesting that the values
in Table \ref{T2} are almost equal to the values of the parameters
$\delta(y)$ and $\eta(y)$ for the rare leptonic decays of the
charged pions $\pi^{\pm} \to e^{\pm} N \to e^{\pm} e^{\pm} \mu^{\mp} \nu$,
Ref.~\cite{CKZ}. The uncertainties in the present Table are in general
smaller, though, because of the high statistics applied in Monte Carlo 
calculations which practically eliminates the numerical uncertainty part.

The rare LNV semihadronic decay widths 
of $M^{\pm}$, cf.~$S_{+}(M)$ of Eq.~(\ref{Spl}),
at first sight  appear to be quartic in the heavy-light
mixing elements $|B_{\ell N}|$ and thus very suppressed.
However, they are proportional to the expressions $\bG(DD^{*})_{jj}$,
Eq.~(\ref{GDD}), which are proportional to $1/\Gamma_{N_j}$
due to the on-shellness of the intermediate $N_j$'s
[cf.~also Eq.~(\ref{P1P1})]. This $1/\Gamma_{N_j}$
is proportional to $1/\K_j \sim 1/|B_{\ell N_j}|^2$ according to
Eqs.~(\ref{GNwidth})-(\ref{calK}). Hence this on-shellness of $N_j$'s
makes these rare process decay widths significantly less suppressed
\be
\label{onsh}
\bG(DD^{*})_{jj} \propto 1/\Gamma_{N_j} \propto 1/\K_j
\propto 1/|B_{\ell N_j}|^2  
\; \Rightarrow \;
S_{+}(M)  \propto  |B_{\ell N_j}|^2 \ .
\ee
However, the expressions (\ref{ImG12}), which appear
in the CP-violating decay width difference $S_{-}(M)$ (\ref{Smi}),
are suppressed by mixings as $\sim |B_{\ell N}|^4$. This means that
in general  $S_{-}(M)$ is
much smaller than the decay width $S_{+}(M) \propto |B_{\ell N_j}|^2$.
Nonetheless, Eqs.~(\ref{ImG12}) show that $S_{-}(M)$
is proportional to $1/\Delta M_N$, and it is this aspect that
represents the opportunity to detect appreciable CP violation
in such decays when $\Delta M_N$ is sufficiently small.
While in general we expect $\Delta M_N \gg \Gamma_{N_j}$,
there exists a well-motivated model \cite{nuMSM,Shapo,nuMSMrev}
with two sterile almost degenerate
neutrinos (where the relation $\Delta M_N \not\gg \Gamma_{N_j}$ is possible) 
in the mass range 
$0.1 \ {\rm GeV} \alt M_{N_j} \alt 10^1 \ {\rm GeV}$.
Our calculations thus suggest that in such a model the
CP violation effects may be appreciable, namely for
$\Delta M_N \sim \Gamma_N$ we obtain $S_{-}(M) \sim S_{+}(M)$
and thus ${\cal A}_{\rm CP}(M) \sim 1$.

For these reasons, from now on we consider the case of near degeneracy:
$\Delta M_N \not\gg \Gamma_N$ (i.e., $\Delta M_N \sim \Gamma_N$). In this
case, several formulas written by now in this Section get even more simplified,
in particular the expressions (\ref{GDD}), (\ref{delX}), (\ref{ImG12}).
Namely, they can be written in terms of the common canonical 
decay width ${\overline S}$
ratio
\ba
{\overline S}(x; x_{\ell_1}, x_{\ell_2}, x') & \equiv &
\frac{3 \pi}{4} \frac{K^2 M_M}{G_F^2}
\frac{1}{x^2}  \lambda^{1/2}(1, x, x_{\ell_1})
\lambda^{1/2}\left( 1, \frac{x'}{x}, \frac{x_{\ell_2}}{x} \right)
Q(x; x_{\ell_1}, x_{\ell_2}, x') \ ,
\label{bBr}
\ea
where we use the notations (\ref{notGDD}) and 
\be
 x \equiv \frac{M_N^2}{M_M^2} \equiv x_2 \approx x_1 \ ,
\label{notx}
\ee
where we denoted by $M_N \equiv M_{N_2} \approx M_{N_1}$. The
function $Q$ is the same as in Eqs.~(\ref{GDD}) and (\ref{ImG12}),
and is given explicitly in Appendix \ref{app3}. In practice we
will need two variants of this function ${\overline S}$,
namely the one for the $DD^{*}$ contributions (${\overline S}^{(D)}$)
and the one of the $CC^{*}$ contributions (${\overline S}^{(C)}$)
\bes
\label{BrDBrC}
\bea
{\overline S}^{(D)}(x) 
& \equiv & {\overline S}(x; x_{\ell_1}, x_{\ell_2}, x') \ ,
\label{BrD}
\\
{\overline S}^{(C)}(x) 
& \equiv & {\overline S}(x; x_{\ell_2}, x_{\ell_1}, x') \ .
\label{BrC}
\eea
\ees
When $\ell_1=\ell_2$ (e.g., when both final leptons are electrons; or both
are muons), the two functions ${\overline S}^{(D)}$ and
${\overline S}^{(C)}$ coincide.
It is straightforward to check that the expressions
of Eqs.~(\ref{GDD}), (\ref{delX}), (\ref{ImG12}) can then be rewritten
in the considered case of nearly degenerate $N_1$ and $N_2$ in terms
of these common functions ${\overline S}^{(X)}$ 
($X=D, C$) and of the heavy-light mixing 
expressions $\K_j$ ($\sim |B_{\ell N_j}|^2$) of Eq.~(\ref{calK})
\bes
\label{Gnor}
\ba
\bG(DD^{*})_{jj}
&=&
\frac{1}{\K_j} {\overline S}^{(D)}(x) \ ,
\qquad
\bG(CC^{*})_{jj} =
\frac{1}{\K_j} {\overline S}^{(C)}(x) \ ,
\label{Gjjnor}
\\
{\rm Re} \bG(DD^{*})_{12} &=&
\delta(y) \frac{2}{(\K_1+\K_2)} {\overline S}^{(D)}(x) \ ,
\quad
{\rm Re} \bG(CC^{*})_{12} =
\delta(y) \frac{2}{(\K_1+\K_2)} {\overline S}^{(C)}(x) \ ,
\label{ReG12nor}
\\
{\rm Im} \bG(DD^{*})_{12} &=&
\frac{\eta(y)}{y} \frac{2}{(\K_1+\K_2)} {\overline S}^{(D)}(x) \ ,
\quad
{\rm Im} \bG(CC^{*})_{12} =
\frac{\eta(y)}{y} \frac{2}{(\K_1+\K_2)} {\overline S}^{(C)}(x) \ ,
\label{ImG12nor}
\ea
\ees
where the definition $y \equiv \Delta M_N/\Gamma_N$ is kept. 

After some straighforward algebra, we can rewrite the sum and difference
$S_{\pm}(M)$ of decay widths, Eqs.~(\ref{Brdef}),
as expressions proportional to these
canonical decay widths ${\overline S}^{(X)}$ ($X=D, C$).
The proportionality factors involve the heavy-light mixing
factors $|B_{\ell N_j}|$ and $\K_j$ [cf.~Eq.~(\ref{calK})], and the
overlap functions $\delta(y)$ and $\eta(y)/y$ tabulated in Table \ref{T2}.
The resulting expressions are
\bes
\label{Brplmi}
\bea
\lefteqn{
S_{+}(M) \equiv
\Gamma(M^- \to \ell_1^- \ell_2^- M^{' +}) + \Gamma(M^+ \to \ell_1^+ \ell_2^+ M^{' -})
}
\nonumber\\
&=&  2 (2 - \delta_{\ell_1 \ell_2})
\left[ \sum_{j=1}^2
\frac{|B_{\ell_1 N_j}|^2 |B_{\ell_2 N_j}|^2}{\K_j}
+ 4 \delta(y) 
\frac{|B_{\ell_1 N_1}||B_{\ell_2 N_1}||B_{\ell_1 N_2}||B_{\ell_2 N_2}|}{(\K_1+\K_2)}
\cos \theta_{21} \right]
\left( {\overline S}^{(D)}(x) +{\overline S}^{(C)}(x) \right),
\label{Brpl}
\\
\lefteqn{
S_{-}(M) \equiv 
 \Gamma(M^- \to \ell_1^- \ell_2^- M^{' +}) - \Gamma(M^+ \to \ell_1^+ \ell_2^+ M^{' -})
}
\nonumber\\
&=&  8 (2 - \delta_{\ell_1 \ell_2}) 
\frac{|B_{\ell_1 N_1}||B_{\ell_2 N_1}||B_{\ell_1 N_2}||B_{\ell_2 N_2}|}{(\K_1+\K_2)}
\; \sin \theta_{21} 
\; \frac{\eta(y)}{y}
\left( {\overline S}^{(D)}(x) +{\overline S}^{(C)}(x) \right) \ .
\label{Brmi}
\eea
\ees
The resulting CP violation ratio ${\cal A}_{\rm CP}(M)$, Eq.~(\ref{Adef}),
can then be written in a form involving only the 
heavy-light mixing factors $|B_{\ell N_j}|$ and $\K_j$ 
[cf.~Eq.~(\ref{calK})], and the
overlap functions $\delta(y)$ and $\eta(y)/y$ tabulated in Table \ref{T2}
\bes
\label{ACP}
\bea
\lefteqn{
{\cal A}_{\rm CP}(M) \equiv  \frac{S_{-}(M)}{S_{+}(M)} 
\equiv  
\frac{ \Gamma(M^- \to \ell_1^- \ell_2^- M^{' +}) - \Gamma(M^+ \to \ell_1^+ \ell_2^+ M^{' -})}
{ \Gamma(M^- \to \ell_1^- \ell_2^- M^{' +}) + \Gamma(M^+ \to \ell_1^+ \ell_2^+ M^{' -})}
}
\nonumber\\
&=& \frac{ \sin \theta_{21} }
{
\left[ \frac{1}{4} 
\sum_{j=1}^2 
\frac{|B_{\ell_1 N_j}|^2 |B_{\ell_2 N_j}|^2}
{|B_{\ell_1 N_1}||B_{\ell_2 N_1}||B_{\ell_1 N_2}||B_{\ell_2 N_2}|}
\frac{(\K_1+\K_2)}{\K_j} + \delta(y) \cos \theta_{21}
\right]
}
\;  \frac{\eta(y)}{y}
\label{ACP1}
\\
&=& \frac{ \sin \theta_{21} }{
\left\{ 
\frac{1}{4} \left[
\kappa_{\ell_1} \kappa_{\ell_2} \left(1 + \frac{\K_1}{\K_2} \right)
+ \frac{1}{\kappa_{\ell_1} \kappa_{\ell_2}} \left(1 + \frac{\K_2}{\K_1} \right)
\right]  + \delta(y) \cos \theta_{21}
\right\}
}
\;  \frac{\eta(y)}{y} \ .
\label{ACP2}
\eea
\ees
In Eq.~(\ref{ACP2}) we used the notations (\ref{kap}).

When $\ell_1=\ell_2$ ($\equiv \ell$), the formulas (\ref{Brplmi})-(\ref{ACP})
simplify because then ${\overline S}^{(D)} = {\overline S}^{(C)} =
{\overline S}$, and $B_{\ell_1 N_j} = B_{\ell_2 N_j} = B_{\ell N_j}$,
$\kappa_{\ell_1} = \kappa_{\ell_2} = \kappa_{\ell}$
\bes
\label{Brplmiell}
\bea
S_{+}(M) =
&=&  4
\left[ \sum_{j=1}^2
\frac{|B_{\ell N_j}|^4}{\K_j}
+ 4 \delta(y) 
\frac{|B_{\ell N_1}|^2 |B_{\ell N_2}|^2}{(\K_1+\K_2)}
\cos \theta_{21} \right]
{\overline S}(x) \ ,
\label{Brplell}
\\
S_{-}(M) 
&=&  16 \frac{ |B_{\ell N_1}|^2 |B_{\ell N_2}|^2}{(\K_1+\K_2)}
\; \sin \theta_{21} 
\; \frac{\eta(y)}{y}
{\overline S}(x) \ ,
\label{Brmiell}
\\
{\cal A}_{\rm CP}(M)
&=& \frac{ \sin \theta_{21} }{
\left\{ 
\frac{1}{4} \left[
\kappa_{\ell}^2 \left(1 + \frac{\K_1}{\K_2} \right)
+ \frac{1}{\kappa_{\ell}^2} \left(1 + \frac{\K_2}{\K_1} \right)
\right]  + \delta(y) \cos \theta_{21}
\right\}
}
\;  \frac{\eta(y)}{y} \ .
\label{ACP2ell}
\eea
\ees

From these expressions and Table \ref{T2} we can deduce:
\begin{enumerate}
\item 
When $y$ becomes large ($y > 10$, i.e., $\Delta M_N > 10 \Gamma_N$),
the CP asymmetries (\ref{Brmi})-(\ref{ACP}) become suppressed
by the small $\eta(y)/y$ factor.
\item
When $y$ is smaller ($y < 10$, i.e., $\Gamma_N < \Delta M_N < 10 \Gamma_N$), 
then the factor $\eta(y)/y$ is comparable with unity, the expressions
$S_{\pm}(M)$ become $\sim  |B_{\ell N_j}|^2 {\overline S}^{(D)}(x)$
(where $x \equiv M_N^2/M_M^2$; $\ell=e, \mu$; 
note that $\K_j \sim |B_{\ell N_j}|^2$);
and the CP violation ratio ${\cal A}_{\rm CP}(M)$ becomes $\sim 1$.
\end{enumerate}

We present in Fig.~\ref{etadelfig} the numerical results of Table
\ref{T2} for the suppression factor $\eta(y)/y$ and for the
overlap factor $\delta(y)$ as a function of
$y \equiv \Delta M_N/\Gamma_N$.
\begin{figure}[htb]
\centering\includegraphics[width=100mm]{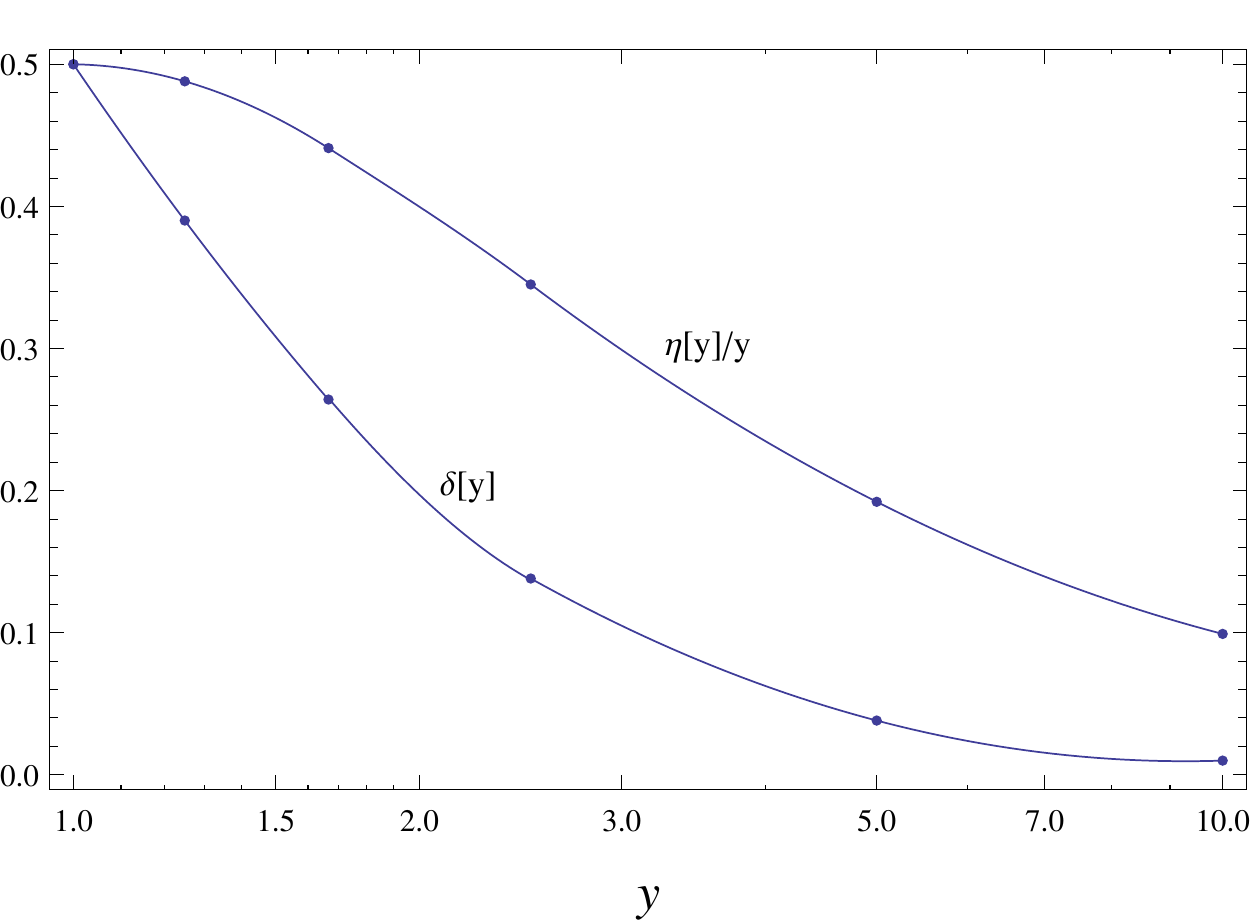}
\vspace{-0.4cm}
\caption{The suppression factors $\eta(y)/y$ and $\delta(y)$, due to the
overlap of the propagator ``resonances'' of $N_1$ and $N_2$, 
as a function of $y \equiv \Delta M_N/\Gamma_N$,
for $1 < y < 10$.}
\label{etadelfig}
\end{figure}

In Ref.~\cite{CDKK}, the decay widths for these processes,
in the case of one (on-shell) neutrino $N$,
$\Gamma(M^{+}) \equiv \Gamma(M^{+} \to \ell^{+} \ell^{+} M^{' -})$, 
were considered. Since in our case
$S_{+}(M) \approx 2  \Gamma(M^{+})$,\footnote{
when neglecting the $N_1$-$N_2$ overlap
effects $\propto \delta(y)$ in $S_{+}(M)$}
the conclusions in Ref.~\cite{CDKK} on the size and
measurability of $\Gamma(M^{+})$ can be  taken over 
as the conclusions on the size and measurability of $S_{+}(M)$ here.
If, in addition, $\Delta M_N \not\gg \Gamma_N$ 
(say: $y \equiv \Delta M_N/\Gamma_N < 5$), these conclusions 
are valid also for the measurability of the
CP-violating decay width difference $S_{-}(M)$ provided
that the phase difference $|\theta_{21}| \sim 1$.\footnote{
We recall that if $y < 5$, we have ${\cal A}_{\rm CP}(M) \sim 1$ and
thus $S_{-}(M) \sim S_{+}(M)$.}

\section{The acceptance factor in the measurement of the considered decays}
\label{sec:acc}

In experiments which try to detect and investigate the 
LNV decay modes of the mesons $M^{\pm}$, 
the (expected) number $N_M \sim 10^N$ of produced mesons $M^{\pm}$ 
(per year, for example) is known. The value of the
corresponding branching ratios of the
LNV decay modes, ${\rm Br}(M^{\pm} \to \ell_1^{\pm} \ell_2^{\pm} M^{' \mp})
\equiv \Gamma(M^{\pm} \to \ell_1^{\pm} \ell_2^{\pm} M^{' \mp})/\Gamma(M^{\pm} \to {\rm all})$, 
then becomes important.
In principle, if ${\rm Br}(M^{\pm} \to \ell_1^{\pm} \ell_2^{\pm} M^{' \mp})
> 10^{-N}$, then such decay modes could be detected. Further,
if an experiment produces approximately equal numbers of
$M^+$ and  $M^-$ mesons, then the branching ratios of
experimental significance for the LNV decays 
$M^{\pm} \to \ell_1^{\pm} \ell_2^{\pm} M^{' \mp}$ are
\bes
\label{Brplmidef}
\bea
{\rm Br}(M) &\equiv& \frac{S_{+}(M)}{\left[\Gamma(M^- \to {\rm all}) + \Gamma(M^+ \to {\rm all}) \right]} \approx \frac{S_{+}(M)}{2 \Gamma(M^- \to {\rm all})} \ ,
\label{Brpldef}
\\
{\cal A}_{\rm CP}(M) {\rm Br}(M) & = & \frac{S_{-}(M)}{\left[\Gamma(M^- \to {\rm all}) + \Gamma(M^+ \to {\rm all}) \right]}  \approx \frac{S_{-}(M)}{2 \Gamma(M^- \to {\rm all})} \ ,
\label{Brmidef}
\eea
\ees
where we use the notation of Eqs.~(\ref{Brdef})-(\ref{Adef}) and (\ref{not0}).
We also used the fact that in the considered cases of pseudoscalar
mesons $M^{\pm}$ the total decay widths $\Gamma(M^- \to {\rm all})$
and  $\Gamma(M^+ \to {\rm all})$ are practically equal.
${\rm Br}(M)$ represents the average of the branching ratios of $M^+$ and $M^-$ 
for these LNV decays, while ${\cal A}_{\rm CP}(M) {\rm Br}(M)$
is the corresponding branching ratio for the
(CP-violating) difference.
The corresponding canonical branching fraction ${\overline {\rm Br}}(M)$
is obtained by dividing the canonical decay width (\ref{bBr}) by
$2 \Gamma(M^- \to {\rm all})$
\be
{\overline {\rm Br}}(x; x_{\ell_1}, x_{\ell_2}, x') \equiv 
\frac{ {\overline S}(x; x_{\ell_1}, x_{\ell_2}, x')}
{2 \Gamma(M^- \to {\rm all})} =
\frac{3 \pi}{8} \frac{K^2 M_M}{G_F^2 \Gamma(M^- \to {\rm all})}
\frac{1}{x^2}  \lambda^{1/2}(1, x, x_{\ell_1})
\lambda^{1/2}\left( 1, \frac{x'}{x}, \frac{x_{\ell_2}}{x} \right)
Q(x; x_{\ell_1}, x_{\ell_2}, x') \ ,
\label{bBrdef}
\ee
where the notations (\ref{notGDD}) and (\ref{notx}) are used.
We have two variants of this function:
the one for the $DD^{*}$ contributions (${\overline {\rm Br}}^{(D)}$)
and the one of the $CC^{*}$ contributions (${\overline {\rm Br}}^{(C)}$),
which are obtained by dividing by $2 \Gamma(M^- \to {\rm all})$
the expressions ${\overline S}^{(D)}$ and ${\overline S}^{(C)}$
of Eqs.~(\ref{BrDBrC}), respectively. When $\ell_1=\ell_2$,
the two functions ${\overline {\rm Br}}^{(D)}$ and
${\overline {\rm Br}}^{(C)}$ coincide ($\equiv {\overline {\rm Br}}$).

Nonetheless, in experiments we must also take into account 
the acceptance (suppression) factor in the detection of these decays, 
which appears due to the small length of
the detector in comparison to the relatively large
lifetime of the (on-shell) sterile neutrinos $N_j$.
Stated otherwise, most of the on-shell neutrinos, produced in the
decay $M^{\pm} \to \ell_1^{\pm} N_j$,
are expected to survive long enough time to travel 
through the detector and decay (into $\ell_2^{\pm} M^{' \mp}$) outside the
detector.\footnote{Only when $M=B$ or $B_c$, a large part of the
produced neutrinos $N_j$ can decay within the detector (see the arguments
later on).} 
This effect suppresses the number of detected decays
and should be taken into account, 
cf.~Refs.~\cite{CDK,scatt3,CKZ,CERN-SPS,commKim}.
The acceptance (suppression) factor is the probability of the 
on-shell neutrino $N$ to decay inside the detector of length $L$
\be
P_{N_j} \approx \frac{L}{\gamma_{N_j} \tau_{N_j} \beta_{N_j}} 
\sim \frac{L}{\gamma_{N_j} \tau_{N_j}}
= \frac{L \Gamma_{N_j}}{\gamma_{N_j}} = 
\frac{L \bG(M_{N_j})}{\gamma_{N_j}} \K_j \equiv  
{\overline A}(M_{N_j}) \K_j \ ,
\label{PN}
\ee
where $\gamma_{N_j}$ is the time dilation (Lorentz) factor
$\gamma_{N_j} = (1 - \beta_{N_j}^2)^{-1/2}$ ($\sim 1$-$10$)
in the lab system.
We took into account that the speed of neutrino is
$\beta_{N_j} \sim 1$. The  quantity $\bG(M_{N_j})$ ($\propto M_{N_j}^5$)
and the factor $\K_j$ ($\propto |B_{\ell N_j}|^2$)
were defined in Eqs.~(\ref{barG}) and (\ref{calK}), respectively.
The quantity ${\overline A}(M_{N_j}) \equiv (L \bG(M_{N_j})/\gamma_{N_j})$
can be called ''canonical acceptance,'' and depends heavily on
the neutrino mass: ${\overline A} \propto M_{N_j}^5$. In Fig.~\ref{tAfig}
we present the values of this canonical acceptance as a function of
the neutrino mass $M_N$, for the choice $L=1$ m 
($= 5.064 \cdot 10^{15} \ {\rm GeV}^{-1}$) and 
$\gamma_{N}=2 $. 
\begin{figure}[htb]
\centering\includegraphics[width=100mm]{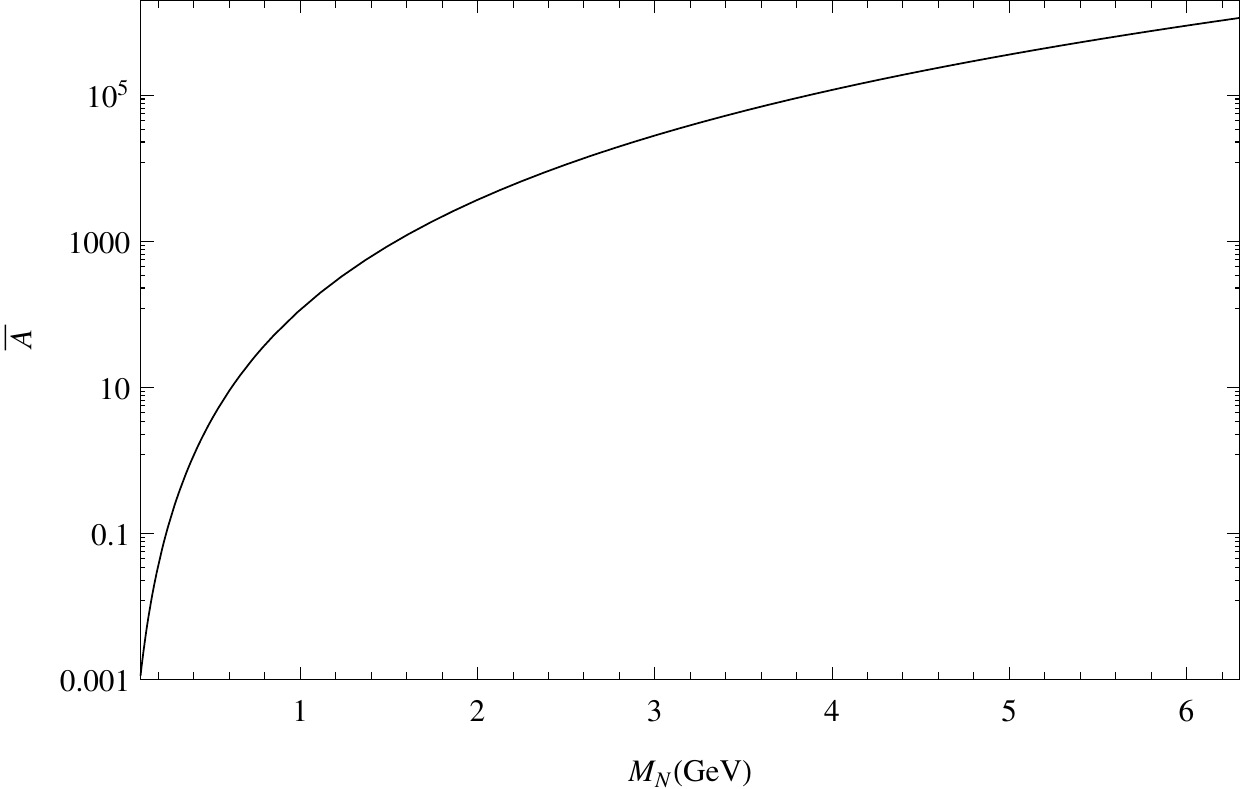}
\vspace{-0.4cm}
\caption{The canonical acceptance  ${\overline A}(M_{N}) \equiv (L \bG(M_{N})/\gamma_{N})$ as a function of the neutrino mass $M_N$. In the curve, 
we took for the length of the
detector the value $L=1$ m and for the time dilation factor 
the value $\gamma_{N}=2$.}
\label{tAfig}
\end{figure}
The values of ${\overline A}$ for other
cases of the values of $L$ and $\gamma_{N}$  are obtained directly
from the presented curve by taking into account that
${\overline A} \propto L/\gamma_{N}$.
The realistic acceptance factor is  then obtained by Eq.~(\ref{PN}), where
$\K_j \sim |B_{\ell N_j}|^2$ ($j=1,2$) 
are the heavy-light mixing factors defined in Eq.~(\ref{calK}) with
coefficients ${\cal N}_{\ell N}$ there of $\sim 10$ according to Fig.~\ref{FigcNellN}.
Combining the results of Fig.~\ref{FigcNellN} with Eq.~(\ref{calK}), we
can write rough approximations for $\K_j$
\bes
\label{calKappr}
\bea
\K_j &\approx& 15 |B_{e N_j}|^2 + 8 |B_{\mu N_j}|^2 + 2  |B_{\tau N_j}|^2 
\quad (K \; {\rm decays}) \ ,
\label{calKapprK}
\\
\K_j &\approx& 7  (|B_{e N_j}|^2 + |B_{\mu N_j}|^2) + 2  |B_{\tau N_j}|^2 \quad (D, D_s \; {\rm decays}) \ ,
\label{calKapprD}
\\
\K_j &\approx& 8  (|B_{e N_j}|^2 + |B_{\mu N_j}|^2) + 3  |B_{\tau N_j}|^2 \quad (B, B_c \; {\rm decays}) \ .
\label{calKapprB}
\eea
\ees
The rough upper bounds for $|B_{\ell N}|^2$, for $\ell = e, \mu, \tau$,
are given in Table \ref{T3} for the typical ranges of our
interest: $M_N$ around $0.25$; $1$; $3$ GeV -- relevant for the decays of
$K$; ($D, D_s$); ($B, B_c$), respectively (see also Table \ref{T1}
for several specific values of $M_N$). 
\begin{table}
\caption{Present rough upper bounds for $|B_{\ell N}|^2$
($\ell = e, \mu, \tau$) for  $M_N$ in the ranges around the values
$0.25$, $1$, $3$ GeV; and
the canonical acceptance factor ${\overline A}(M_N)$ (for $L=1$ m and
$\gamma_N=2$).}
\label{T3}
\begin{tabular}{|l|lll|l|}
$M_N$ [GeV] & $|B_{e N}|^2$ &  $|B_{\mu N}|^2$ & $|B_{\tau N}|^2$ & ${\overline A}$
\\
\hline
$\approx 0.25$ & $10^{-8}$ & $10^{-7}$ & $10^{-4}$ & 0.11
\\
$\approx 1.0$ &  $10^{-7}$ & $10^{-7}$ & $10^{-2}$ & 115.
\\
$\approx 3.0$ &  $10^{-6}$ & $10^{-4}$ & $10^{-4}$ & $3 \cdot 10^4$
\end{tabular}
\end{table}
The corresponding values of the
canonical acceptance factor ${\overline A}(M_N)$ are also included.
Combining Eqs.~(\ref{PN}) with (\ref{calKappr}) and Table \ref{T3},
we obtain for the acceptance factor $P_{N_j}$ the following
estimates and upper bounds relevant for the $K$  decays ($M_N \approx 0.25$ GeV),
$D$ and $D_s$ decays ($M_N \approx 1$ GeV), and $B$ and $B_c$ decays
($M_N \approx 3$ GeV):
\bes
\label{uppPN}
\bea
P_{N_j}(M_N\approx 0.25{\rm GeV}) & \approx & 
1.7 |B_{e N_j}|^2 + 0.9 |B_{\mu N_j}|^2  \quad (+  0.2 |B_{\tau N_j}|^2)
\nonumber\\
&\alt& 10^{-8} + 10^{-7} \quad (+ 10^{-5}) \ ,
\label{uppPNK}
\\
P_{N_j}(M_N\approx 1{\rm GeV}) & \approx & 
0.8 \cdot 10^3 |B_{e N_j}|^2 + 0.8 \cdot 10^3 |B_{\mu N_j}|^2 
\quad (+  2 \cdot 10^2 |B_{\tau N_j}|^2)
\nonumber\\
&\alt& 10^{-4} + 10^{-4} \quad (+ 10^{0}) \ ,
\label{uppPND}
\\
P_{N_j}(M_N\approx 3{\rm GeV}) & \approx & 
3 \cdot 10^5 |B_{e N_j}|^2 + 3 \cdot 10^5 |B_{\mu N_j}|^2 
\quad (+  1 \cdot 10^5 |B_{\tau N_j}|^2)
\nonumber\\
&\alt& 10^{0} + 10^{0} \quad (+ 10^{0}) \ ,
\label{uppPNB}
\eea
\ees
The upper bounds for $P_{N_j}$ in Eqs.~(\ref{uppPN}) are written as a sum
of the contributions of upper bounds from $|B_{e N_j}|^2$, $|B_{\mu N_j}|^2$
and  $|B_{\tau N_j}|^2$ separately. Further, the contributions of
$|B_{\tau N_j}|^2$ are included in Eqs.~(\ref{uppPN}) optionally, 
in the parentheses, because the upper bounds
of the mixings $|B_{\tau N_j}|^2$ are still very high and are
expected to be reduced significantly in the foreseeable future.
The upper bounds which give results higher than one are replaced by
one ($10^0$), because the acceptance (decay probability) $P_{N_j}$
can never be higher than one by definition.

From now on in this Section, we will assume the following:
\bes
\label{assum}
\bea
|B_{\ell N_1}|^2 &\sim & |B_{\ell N_2}|^2  \equiv |B_{\ell N}|^2
\label{assum1}
\\ 
\Rightarrow \ \K_1 &\sim& \K_2 \equiv \K \  .
\label{assum2}
\eea
\ees
In addition, we consider that it is the flavor $\ell$ which has the
dominant (largest) mixing $|B_{\ell N}|^2$. Then we have
\be
\K \approx {\cal N}_{\ell N} |B_{\ell N}|^2 \sim 10 \; |B_{\ell N}|^2 \ .
\label{Kappr}
\ee
The dominant branching ratios
${\rm Br}(M)$ and ${\cal A}_{\rm CP}(M){\rm Br}(M)$ 
will then be, according to the obtained
expressions (\ref{Brplmi}) and (\ref{Brplmiell})
[together with the definitions (\ref{Brplmidef})-(\ref{bBrdef})], 
those which have in the final
state two equal charged leptons $\ell$ with dominant mixing:
$M^{\pm} \to \ell^{\pm} \ell^{\pm} M^{' \mp}$.

The theoretical branching ratios 
${\rm Br}(M)$ and ${\cal A}_{\rm CP}(M) {\rm Br}(M)$, 
Eqs.~(\ref{Brplmidef}), can be obtained by diividing
Eqs.~(\ref{Brplell})-(\ref{Brmiell}) by
$2 \Gamma(M^- \to {\rm all})$. Using in addition 
Eqs.~(\ref{assum})-(\ref{Kappr}) and the definition
(\ref{bBrdef}), this gives
\bes
\label{Brth}
\bea
{\rm Br}(M) & \sim & 8 \frac{|B_{\ell N}|^4}{\K} {\overline {\rm Br}}(x)
\sim  {\overline {\rm Br}}(x) |B_{\ell N}|^2 \ ,
\label{Brthpl}
\\
{\cal A}_{\rm CP}(M) {\rm Br}(M) & \sim & 8 \frac{|B_{\ell N}|^4}{\K} 
\sin \theta_{21} \frac{\eta(y)}{y} {\overline {\rm Br}}(x)
\sim  {\overline {\rm Br}}(x) |B_{\ell N}|^2 \sin \theta_{21} \ ,
\label{Brthmi}
\eea
\ees
where in the last relation we took into account that $\eta(y)/y \sim 1$
(since $\Delta M_N \not\gg \Gamma_N$ in our considered cases).

The effective (i.e., experimental) branching ratios 
${\rm Br}^{\rm (eff)}(M) = P_N {\rm Br}(M)$ and
${\cal A}_{\rm CP}(M){\rm Br}^{\rm (eff)}(M)$
can be estimated,
in the considered case of Eqs.~(\ref{assum})-(\ref{Kappr}),
in the following way
[using Eqs.~(\ref{PN}) and (\ref{Brth})]:
\bes
\label{PNBr}
\bea
{\rm Br}^{\rm (eff)}(M) &\equiv &P_N {\rm Br}(M) \sim {\overline A}(M_N) \K {\rm Br}(M) \sim
{\overline A}(M_N) \K \left( \frac{8 |B_{\ell N}|^4}{\K} {\overline {\rm Br}}(x)
\right) 
\nonumber\\
&=& 
\left[ 8 {\overline A}(M_N) {\overline {\rm Br}}(x) \right] |B_{\ell N}|^4 \ ,
\label{PNBrpl}
\\
{\cal A}_{\rm CP}(M) {\rm Br}^{\rm (eff)}(M) &\equiv& P_N {\cal A}_{\rm CP}(M) {\rm Br}(M) \sim {\overline A}(M_N) \K {\rm Br}_{-}(M) \sim
{\overline A}(M_N) \K \left( \frac{8 |B_{\ell N}|^4}{\K} 
\sin \theta_{21} \frac{\eta(y)}{y}  {\overline {\rm Br}}(x) \right) 
\nonumber\\
&=&
8 {\overline A}(M_N) |B_{\ell N}|^4 \sin \theta_{21} \frac{\eta(y)}{y}  
{\overline {\rm Br}}(x)
\sim 
\left[ 8 {\overline A}(M_N)  {\overline {\rm Br}}(x) \right]
|B_{\ell N}|^4 \sin \theta_{21} \ ,
\label{PNBrmi}
\eea
\ees
where in the last line of Eq.~(\ref{PNBrmi}) we took into account that
$\eta(y)/y \sim 1$ (true when $\Delta M_N \not\gg \Gamma_N$). Furthermore,
since $\ell_1=\ell_2=\ell$ in the considered case, the 
canonical branching fractions are equal: ${\overline {\rm Br}}^{(C)}(x)=
 {\overline {\rm Br}}^{(D)}(x) \equiv  {\overline {\rm Br}}(x)$;
and we recall that $x \equiv (M_N/M_M)^2$. We see that in Eqs.~(\ref{PNBr}) the
most important factor at $|B_{\ell N}|^4$ is the ``effective'' canonical
branching ratio
\be
{\overline {\rm Br}}_{\rm eff}(M_N) \equiv 8  {\overline A}(M_N) {\overline {\rm Br}}(x) \ .
\label{bBReff}
\ee
Only in the case of $B^{\pm}$ and $B_c^{\pm}$ LNV decays 
we could have $P_N \sim 1$,
Eq.~(\ref{uppPNB}), and in such a case Eqs.~(\ref{PNBr}) do not apply,
but rather Eqs.~(\ref{Brth}).
In Figs.~\ref{brKfig}-\ref{brBcfig} we present the
effective canonical branching ratios (\ref{bBReff}) as a function of the
neutrino mass $M_N$, for various considered LNV decays
of the type $M^{\pm} \to \ell^{\pm} \ell^{\pm} M^{' \mp}$, where:
$M=K$ in Fig.~\ref{brKfig}; $M=D, D_s$ in Figs.~\ref{brDDsfig}(a), (b);
$M=B, B_c$ in Figs.~\ref{brBfig}(a) and \ref{brBcfig}(a), respectively.
In general $\ell=e, \mu$. We took $L=1$ m and $\gamma_N=2$.
In addition, for the case when $P_N \sim 1$ and 
consequently the estimates Eqs.~(\ref{Brth}) apply, we present
in  Figs.~\ref{brBfig}(b) and
\ref{brBcfig}(b) the theoretical branching ratios
${\overline {\rm Br}}(x)$ as a function of $M_N$ for $B^{\pm}$ and
$B_c^{\pm}$ decays, respectively.\footnote{
Our formulas permit also evaluation of ${\overline {\rm Br}}_{\rm eff}$
and  ${\overline {\rm Br}}(x)$ for the decays 
$M^{\pm} \to \ell_1^{\pm} \ell_2^{\pm} M^{' \mp}$ when $\ell_1 \not= \ell_2$.
And also when the final leptons
are $\tau$ leptons (and $M^{\pm} = B^{\pm}$ or $B_c^{\pm}$), with
the values similar to those in Figs.~\ref{brBfig} and \ref{brBcfig},
except that the range of $M_N$ is now significantly shorter:
$M_{M^{'}} + M_{\tau} < M_N < M_M - M_{\tau}$.}
For the CKM matrix elements and the meson decay constants, appearing
in $K^2$ factor defined in Eq.~(\ref{Ksqr}), and for masses and
lifetimes of the mesons, we used the values
of Ref.~\cite{PDG2012}; and for the decay constants $f_B$ and $f_{B_c}$
we used the values of Ref.~\cite{CKWN}: $f_B = 0.196$ GeV,
$f_{B_c}=0.322$ GeV.
\begin{figure}[htb]
\centering\includegraphics[width=90mm]{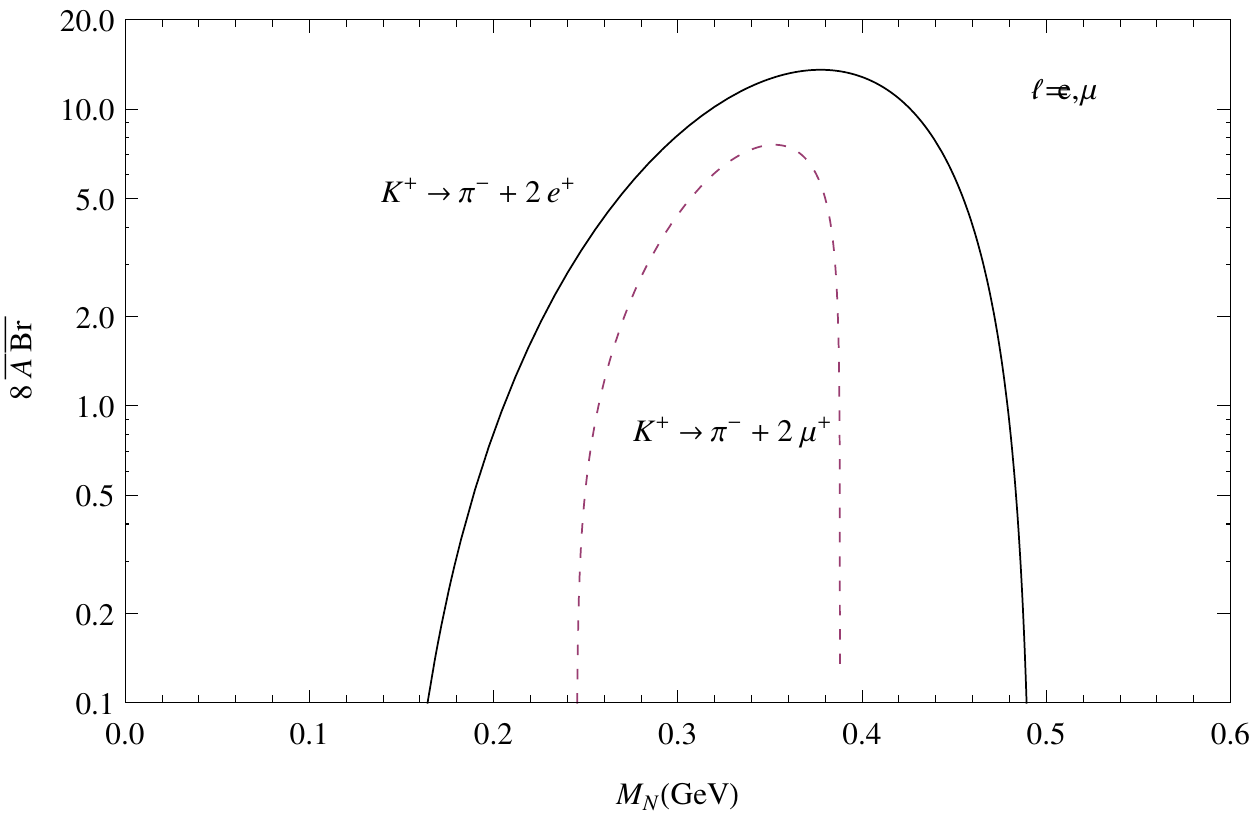}
\vspace{-0.4cm}
\caption{The effective canonical branching ratio (\ref{bBReff})
for the  $K^{\pm} \to \ell^{\pm} \ell^{\pm} \pi^{' \mp}$ decays
($\ell=e, \mu$) as a function of the Majorana neutrino mass $M_N$.}
\label{brKfig}
\end{figure}
\begin{figure}[htb] 
\begin{minipage}[b]{.49\linewidth}
\includegraphics[width=85mm]{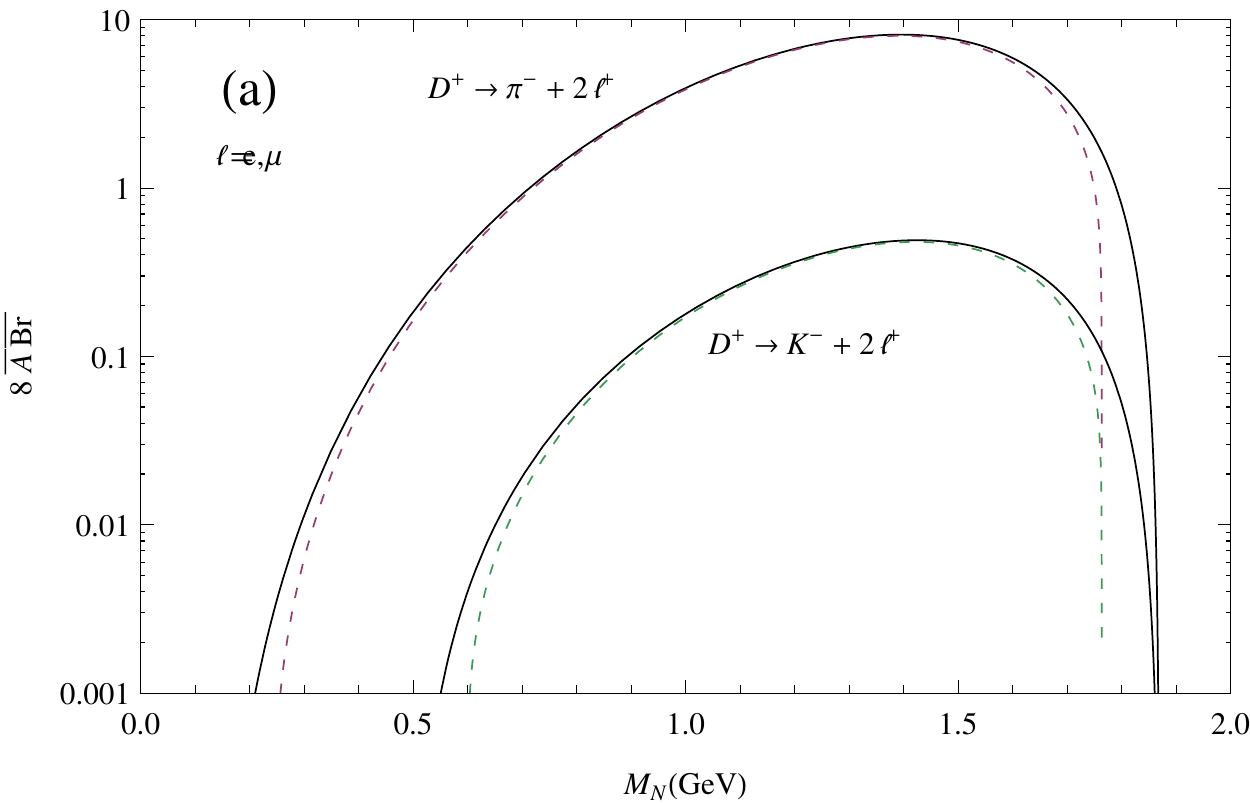}
\end{minipage}
\begin{minipage}[b]{.49\linewidth}
\includegraphics[width=85mm]{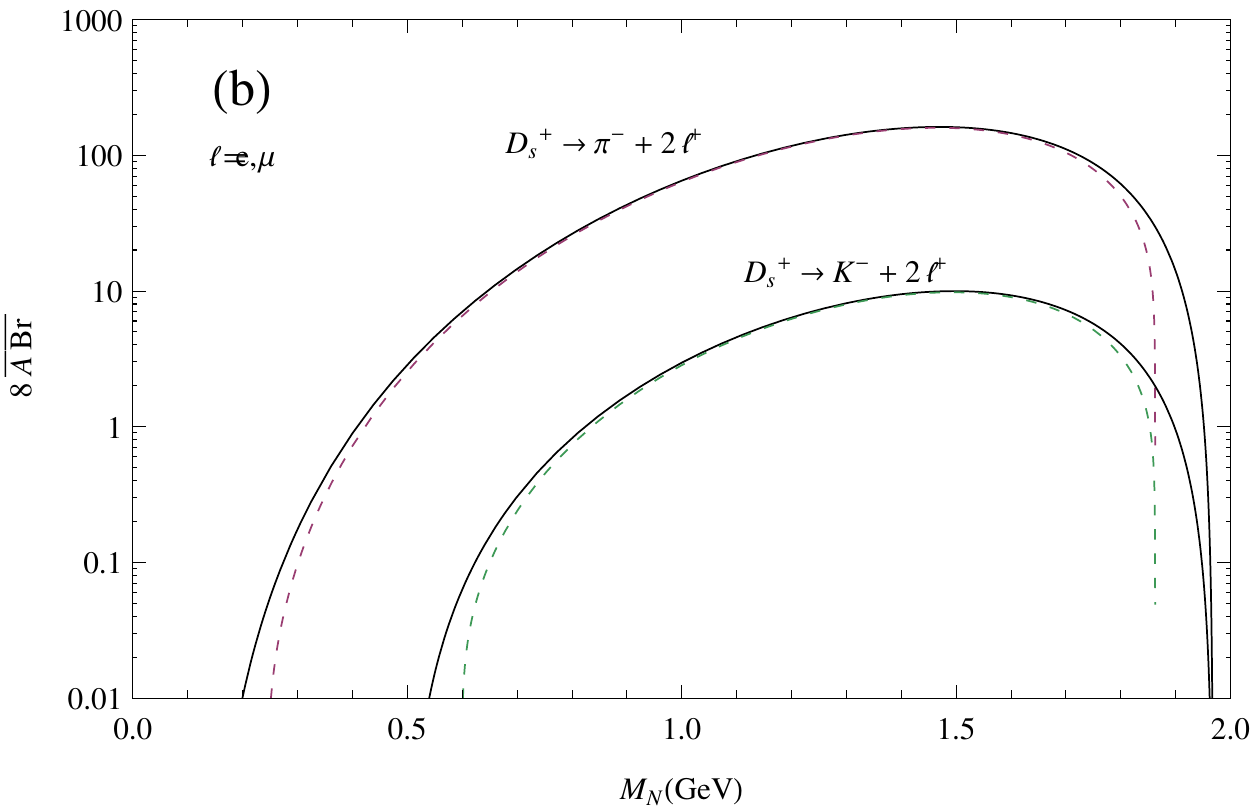}
\end{minipage}
\vspace{-0.4cm}
\caption{The effective canonical branching ratio (\ref{bBReff})
as a function of the Majorana neutrino mass $M_N$
for the  LNV decays of: (a) $D^{\pm}$ mesons; (b) $D_s^{\pm}$ mesons.
The solid lines are for $\ell=e$, and the dashed lines for $\ell=\mu$.}
\label{brDDsfig}
 \end{figure}
\begin{figure}[htb] 
\begin{minipage}[b]{.49\linewidth}%
\includegraphics[width=85mm]{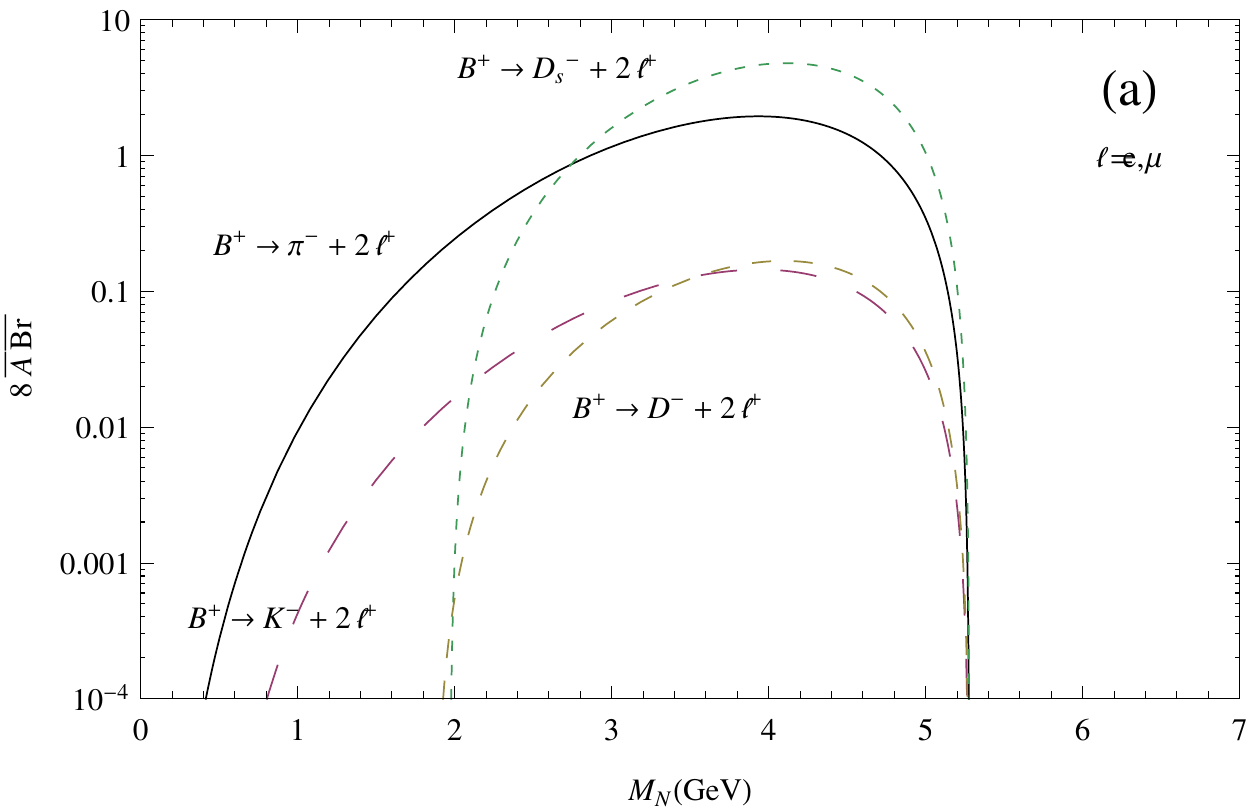}
\end{minipage}
\begin{minipage}[b]{.49\linewidth}
\includegraphics[width=85mm]{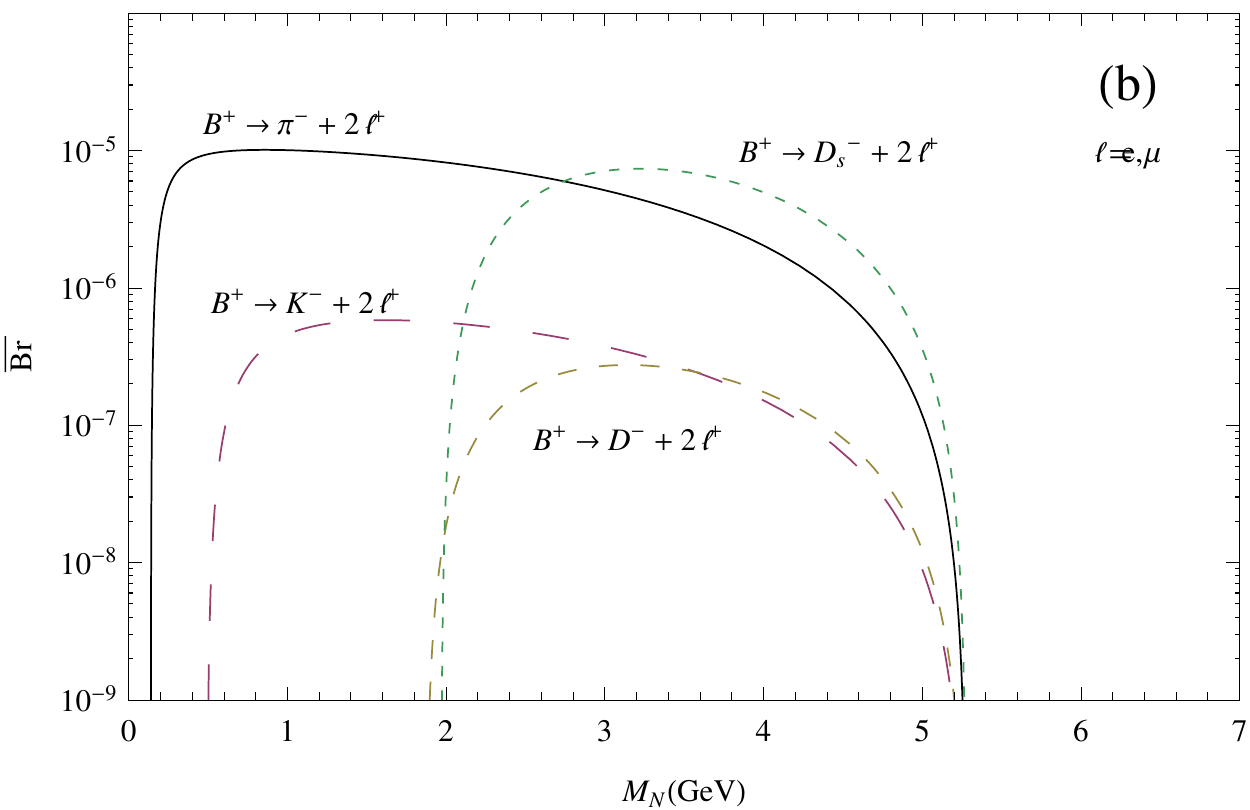}
\end{minipage}
\vspace{-0.4cm}
\caption{(a) The effective canonical branching ratio (\ref{bBReff})
as a function of the Majorana neutrino mass $M_N$
for the  LNV decays of $B^{\pm}$ mesons, 
$B^{\pm} \to \ell^{\pm} \ell^{\pm} M^{' \mp}$,
where $\ell=e, \mu$ (no discernible difference between the two cases);
 (b) the corresponding
curves for the theoretical 
canonical branching ratio ${\overline {\rm Br}}$.}
\label{brBfig}
 \end{figure}
\begin{figure}[htb] 
\begin{minipage}[b]{.49\linewidth}
\includegraphics[width=85mm]{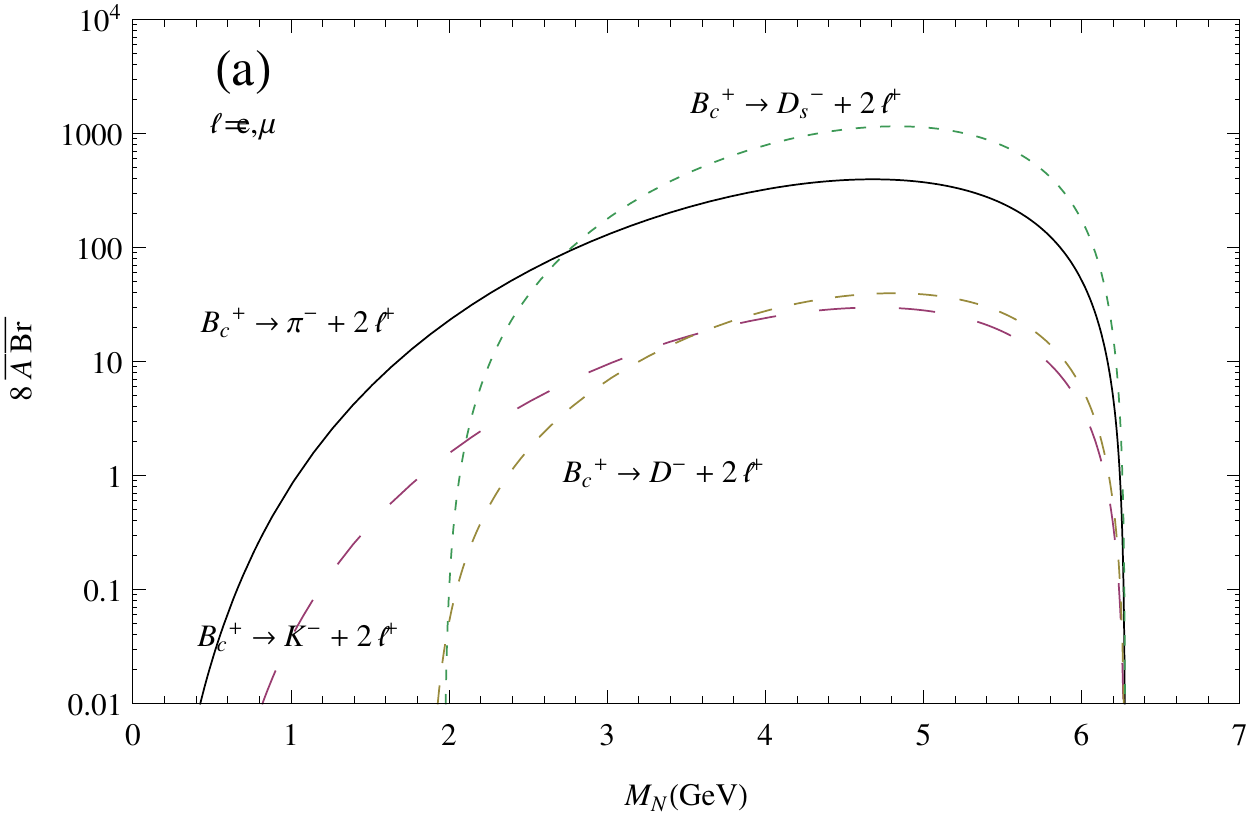}
\end{minipage}
\begin{minipage}[b]{.49\linewidth}
\includegraphics[width=85mm]{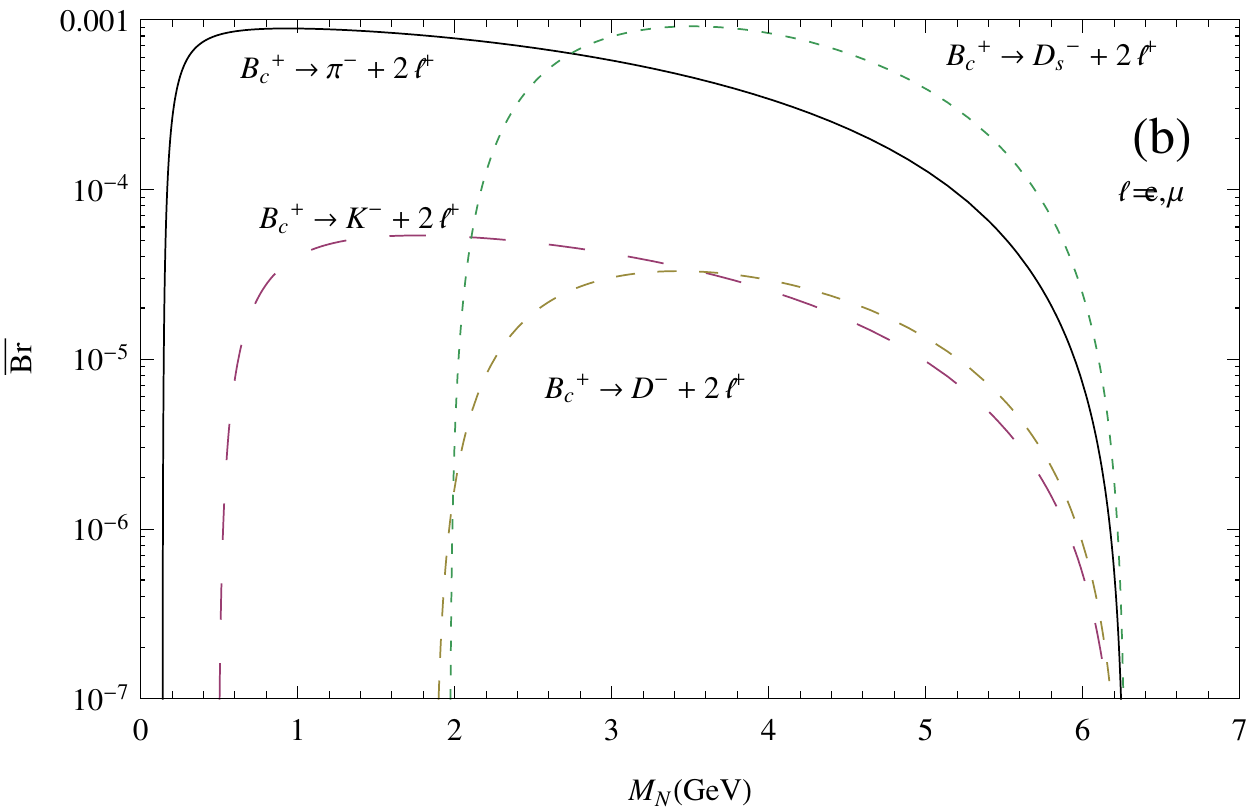}
\end{minipage}
\vspace{-0.4cm}
\caption{The same as in Fig.~\ref{brBfig}, but for the LNV decays of
the charmed mesons $B_c^{\pm}$.}
\label{brBcfig}
 \end{figure}

In Table \ref{T4}
we display some values of the factor
${\overline {\rm Br}}_{\rm eff}$, for the 
representative values of $M_N$ in the decays
 $M^{\pm} \to \ell^{\pm} \ell^{\pm} M^{' \mp}$.
\begin{table}
\caption{Values of the factor $8 {\overline A}(M_N) {\overline {\rm Br}}(x)$
(with $L=1$ m and $\gamma_N=2$) for some of the considered LNV decays:
$M^{\pm} \to \ell^{\pm} \ell^{\pm} \pi^{' \mp}$.
We chose $M_N$ such that the maximal value is obtained (this value 
of $M_N$ is given in parentheses, in GeV).
For the $K$ decay, the two different values are
given for $\ell=e$ and $\ell=\mu$. For all other decays $\ell=\mu$ is chosen
(the values for $\ell=e$ are similar).} 
\label{T4}
\begin{tabular}{|l|l|l|l|l|l|l|}
$M^{\pm}$: & $K^{\pm}$ ($\ell=e$)  & $K^{\pm}$ ($\ell=\mu$)  & $D^{\pm}$ & $D_s^{\pm}$ & $B^{\pm}$ & $B_c^{\pm}$
\\
\hline
$8 {\overline A} {\overline {\rm Br}}$: & 13.5 (0.38) & 7.5 (0.35) &
 8. (1.39) & 159. (1.47) &  1.93 (3.9) & 395. (4.7)
\\
\end{tabular}
\end{table}

Let us now take, as an example, 
the decays $D_s^{\pm} \to \mu^{\pm} \mu^{\pm} \pi^{\mp}$,\footnote{
This is one of the preferred decay modes proposed at
CERN-SPS \cite{CERN-SPS}.}
and let us assume that $|B_{\mu N}|^2$ is the dominant mixing (i.e.,
$\ell=\mu$).
Then Eqs.~(\ref{PNBr}) and Table \ref{T4} 
imply that the effective (experimentally measurable)
sum $P_N {\rm Br}(D_s)$ 
and difference $P_N {\cal A}_{\rm CP}(D_s) {\rm Br}(D_s)$
of the branching ratios for
these decays are
\bes
\label{PNBrDs}
\bea
{\rm Br}^{\rm (eff)}(D_s) \equiv P_N {\rm Br}(D_s) &\sim& 10^2 |B_{\mu N}|^4 \ ,
\label{PNBrplDs}
\\
{\cal A}_{\rm CP}(D_s) {\rm Br}^{\rm (eff)}(D_s) \equiv  P_N {\cal A}_{\rm CP}(D_s) {\rm Br}(D_s) & \sim & 10^2 |B_{\ell N}|^4  \sin \theta_{21} \frac{\eta(y)}{y} 
\sim  10^2 |B_{\ell N}|^4  \sin \theta_{21} \ .
\label{PNBrmiDs}
\eea
\ees
Taking into  account that in such decays the present rough upper bound on
the mixing is $|B_{\mu N}|^2 \alt 10^{-7}$ (cf.~Table \ref{T3}),
Eqs.~(\ref{PNBrDs}) imply that $P_N {\rm Br}(D_s) \alt 10^{-12}$.
The proposed experiment at CERN-SPS \cite{CERN-SPS} would produce the numbers
of $D$ and $D_s$ mesons by several orders higher than $10^{12}$ and
would thus be able to explore whether there is a production
of the sterile Majorana neutrinos $N_j$. Furthermore, if there are two
almost degenerate neutrinos (as is the case in the $\nu$MSM model
\cite{nuMSM,Shapo}), then in such a case it is
possible that $y (\equiv \Delta M_N/\Gamma_N)
\not\ll 1$, and thus $\eta(y)/y \sim 1$. Then the 
estimate (\ref{PNBrmiDs}) would imply that the CP-violating
difference of effective branching ratios 
$P_N {\cal A}_{\rm CP}(D_s) {\rm Br}(D_s)$
is of the same order as the sum $P_N {\rm Br}(D_s)$
(provided that the phase difference $|\theta_{21}| \not\ll 1$).
This means that if experiments discover the aforementioned
$\nu$MSM-type Majorana neutrinos, they will possibly discover also
CP violation in the Majorana neutrino sector.

\section{Conclusions}
\label{sec:concl}

We investigated the possibility of detection of CP violation in
lepton number violating (LNV) semihadronic decays
$M^{\pm} \to \ell_1^{\pm} \ell_2^{\pm} M^{' \mp}$,
where $M$ and $M^{'}$ are pseudoscalar mesons,
$M=K, D, D_s, B, B_c$ and $M^{'}=\pi, K, D, D_s$, and the charged
leptons are $\ell_1, \ell_2 = e, \mu$.
The decay widths of such decays,
mediated by on-shell sterile Majorana neutrinos $N$ with masses
$M_N \sim 1$ GeV, have been studied by various authors,
cf.~Refs.~\cite{LittSh,DGKS,Ali,IvKo,GoJe,Atre,HKS,CDKK},
with a view of a possible detection in future experiments
such as the proposed CERN-SPS experiment \cite{CERN-SPS}.
In the present work we investigated the possibility of
detecting the CP-violating decay width difference
$S_{-}(M) \equiv 
[\Gamma(M^- \to \ell_1^- \ell_2^- M^{' +})-\Gamma(M^+ \to \ell_1^+ \ell_2^+ M^{' -})]$ 
in such processes,
in the scenarios of two on-shell sterile Majorana neutrinos $N_1$, $N_2$.
We used the same approach as in our previous work \cite{CKZ}
where CP violation was investigated in purely leptonic rare decays
$\pi^{\pm} \to  e^{\pm} e^{\pm} \mu^{\mp} \nu$: the crucial aspect 
is the expression for the imaginary part of the product of
the propagators of two Majorana neutrinos, Eqs.~(\ref{ImP1P2gen}).
A central point, as in Ref.~\cite{CKZ}, is that when the difference of masses 
$\Delta M_N \equiv M_{N_2} - M_{N_1}$ ($>0$) of the two
sterile neutrinos becomes small enough,
comparable to the (small) total decay widths of these neutrinos,
$\Delta M_N \not\gg \Gamma_N$, the mentioned imaginary part
becomes large and leads to a
large CP-violating decay width difference
$S_{-}(M)$. We show that in such a case, and provided that a specific
CP-violating difference $\theta_{21}$ of the phases of
heavy-light neutrino mixings is not very small
($|\theta_{21}| \not\ll 1$), the decay width difference
$S_{-}(M)$ becomes comparable with the sum of the decay widths
of the LNV decays $S_{+}(M) \equiv 
 [\Gamma(M^- \to \ell_1^- \ell_2^- M^{' +})+\Gamma(M^+ \to \ell_1^+ \ell_2^+ M^{' -})]$,
and the corresponding CP ratio
${\cal A}_{\rm CP}(M) \equiv S_{-}(M)/S_{+}(M)$ thus becomes
$ {\cal A}_{\rm CP}(M) \sim 1$. It is interesting that the
requirement of the near degeneracy of the two sterile neutrinos 
(with $M_{N_j} \sim 1$ GeV), at which we arrive by
requiring appreciable CP violation, fits well into the 
well-motivated $\nu$MSM model
\cite{nuMSM,Shapo,nuMSMrev}, where the near degeneracy of
the two sterile neutrinos with mass $M_{N_j} \sim 1$ GeV is 
obtained by requiring that the third (the lightest) sterile neutrino be
the dark matter candidate. The results of our calculation can thus be
interpreted in the framework of the $\nu$MSM model, namely
that if the model is experimentally confirmed then it is possible
that significant neutrino sector CP violation effects will be detected as well. 

\begin{acknowledgments}
\noindent
This work was supported in part by FONDECYT Grant No.~1130599 (G.C. and C.S.K.), and
by CONICYT Fellowship ``Beca de Doctorado Nacional'' and Proyecto PIIC 2013 (J.Z.S.).
The work of C.S.K. was supported in part by the NRF grant funded by the Korean government of the MEST (No. 2011-0017430) and (No. 2011-0020333).
We thank Marco Drewes for bringing to our attention Ref.~\cite{DM}.
\end{acknowledgments}

\appendix
\section{Explicit formulas for the $M^{\pm} \to \ell_1^{\pm} \ell_2^{\pm} M^{' \mp}$ decay width}
\label{app1}

The matrix element ${\cal T}(M^{\pm})$ for the decay of Fig.~\ref{FigLV} can be
written in the form
\be
{\cal T}(M^{\pm}) =  K_{\pm} \sum_{j=1}^2 k_j^{(\pm)} M_{N_j} \left[
P_j(D) T_{\pm}(D) + P_j(C) T_{\pm}(C) \right] \ ,
\label{cTMpm}
\ee
where $j=1,2$ refer to the contributions of the exchanges of the two
intermediate neutrinos $N_j$, and $X=D, C$ refer to the contribution of the
direct and crossed channels, respectively, cf.~Fig.~\ref{FigLV}. 
In Eq.~(\ref{cTMpm}), $k_j^{(\pm)}$ are the heavy-light mixing factors defined 
in Eq.~(\ref{kj}); $P_j(X)$ ($j=1,2; X=D, C$) are the propagator
functions of $N_j$ neutrino for the $D$ and $C$ channel, Eqs.~(\ref{Pj}),
and $K_{\pm}$ are the constants coming from the vertices
\be
K_{-} = - G_F^2 V_{Q_u Q_d} V_{q_u q_d} f_M f_{M'} \ , \qquad
K_{+} = (K_{-})^{*} \ ,
\label{Kmp}
\ee
where $f_{M}$ and $f_{M'}$ are the decay constants of $M^{\pm}$ and $M^{' \mp}$,
and $V_{Q_u Q_d}$ and $V_{q_u q_d}$ are the CKM elements for $M^{\pm}$ and $M^{' \mp}$:
$M^+$ has the valence quark content 
$Q_u {\bar Q}_d$; $M^{' +}$ has $q_u {\bar q}_d$.
The functions $T_{\pm}(D)$ and $T_{\pm}(C)$ appearing in the amplitude
(\ref{cTMpm}) can be written as
\bes
\label{Tpm}
\bea
T_{\pm}(D) & = & {\overline u}_{\ell_2}(p_2) \pMpsl \pMsl (1 \mp \gamma_5) v_{\ell_1}(p_1) \ ,
\label{TDpm}
\\
T_{\pm}(C) & = & {\overline u}_{\ell_2}(p_2) \pMsl \pMpsl (1 \mp \gamma_5) v_{\ell_1}(p_1) \ ,
\label{TCpm}
 \eea
\ees
where the spinors are written in the helicity basis. Squaring and summing over the
final helicities leads to the square $|{\cal T}(M^{\pm})|^2$
of the total decay amplitude (\ref{cTMpm}) as given in Eq.~(\ref{GM2}) in
conjuntion with Eqs.~(\ref{kj})-(\ref{Ksqr}), where the quadratic expressions
$T_{\pm}(X) T_{\pm}(Y)^{*}$ ($X,Y=D,C$) appearing in the normalized decay
widths $\bG_{\pm}(XY^{*})_{ij}$ in Eq.~(\ref{bGXY}) are
\bes
\label{TT}
\bea
 T_{\pm}(D) T_{\pm}(D)^{*} & = & 8 {\big [} M_M^2 M_{M'}^2 (p_1 \cdot p_2)
- 2 M_M^2 (p_1 \cdot p_{M'}) (p_2 \cdot p_{M'}) -2  M_{M'}^2 (p_1 \cdot p_{M}) (p_2 \cdot p_{M})
\nonumber\\
&&
+ 4 (p_1 \cdot p_{M}) (p_2 \cdot p_{M'}) (p_M \cdot  p_{M'}) {\big ]}
\equiv T(D) T(D)^{*} \ ,
\label{TDTD}
\\
 T_{\pm}(C) T_{\pm}(C)^{*} & = & 8 {\big [} M_M^2 M_{M'}^2 (p_1 \cdot p_2)
- 2 M_M^2 (p_1 \cdot p_{M'}) (p_2 \cdot p_{M'}) -2  M_{M'}^2 (p_1 \cdot p_{M}) (p_2 \cdot p_{M})
\nonumber\\
&&
+ 4 (p_2 \cdot p_{M}) (p_1 \cdot p_{M'}) (p_M \cdot  p_{M'})  {\big ]}
\equiv T(C) T(C)^{*} \ ,
\label{TCTC}
\\
T_{\pm}(D) T_{\pm}(C)^{*} & = & 16 {\Big \{}
M_M^2 (p_1 \cdot p_{M'}) (p_2 \cdot p_{M'})  +  M_{M'}^2 (p_1 \cdot p_{M}) (p_2 \cdot p_{M})
- \frac{1}{2} M_M^2 M_{M'}^2 (p_1 \cdot p_2)
\nonumber\\
&&
+ (p_M \cdot  p_{M'}) 
\left[ 
-(p_1 \cdot p_{M}) (p_2 \cdot p_{M'})
-(p_2 \cdot p_{M}) (p_1 \cdot p_{M'})
+  (p_M \cdot  p_{M'}) (p_1 \cdot p_2) \right]
\nonumber\\
&&
\mp i (p_M \cdot  p_{M'}) \epsilon(p_M,p_1,p_2,p_{M'}) 
{\Big \}} 
\label{TDTC}
\\
T_{\pm}(C) T_{\pm}(D)^{*} & = & \left( T_{\pm}(D) T_{\pm}(C)^{*} \right)^{*}
= T_{\mp}(D) T_{\mp}(C)^{*} = \left(  T_{\mp}(C) T_{\mp}(D)^{*}  \right)^{*} \ ,
\label{TCTD}
\eea
\ees
where in these expressions the summation over the (final) helicities of
the leptons $\ell_1$ and $\ell_2$ is implied, and we denoted
\be
\epsilon(q_1, q_2, q_3, q_4) \equiv \epsilon^{\eta_1 \eta_2 \eta_3 \eta_4}
(q_1)_{\eta_1} (q_2)_{\eta_2} (q_3)_{\eta_3} (q_4)_{\eta_4} \ ,
\label{eps}
\ee
and $\epsilon^{\eta_1 \eta_2 \eta_3 \eta_4}$ is the totally antisymmetric Levi-Civita tensor with the
sign convention $\epsilon^{0123}=+1$.

The expressions (\ref{TT}), in conjunction with the definitions (\ref{bGXY}), imply for
the normalized decay widths $\bG_{\pm}(XY^{*})_{ij}$ of Eq.~(\ref{bGXY})
various symmetry relations, among them that $\bG_{\pm}(DD^{*})$ and
$\bG_{\pm}(CC^{*})$ are both self-adjoint ($2 \times 2$) matrices
and that elements of the $D$-$C$ interference matrices
$\bG_{\pm}(CD^{*})$ and $\bG_{\pm}(DC^{*})$ are related
\bes
\label{symm}
\bea
\bG(DD^{*})_{ij} & = & \left( \bG(DD^{*})_{ji} \right)^{*} \ ,
\qquad
\bG(CC^{*})_{ij} =  \left( \bG(CC^{*})_{ji} \right)^{*} \ ,
\label{symmDD}
\\
\bG_{\pm}(CD^{*})_{ij} & = &  \left( \bG_{\pm}(DC^{*})_{ji} \right)^{*} \ .
\label{symmCD}
\eea
\ees
When the two final leptons are the same ($\ell_1=\ell_2$), we can use the fact
that the integration $d_3$ over the final particles is symmetric under
$(p_1 \leftrightarrow p_2)$ (because $M_{\ell_1} = M_{\ell_2}$), and we have
additional symmetry relations
\bes
\label{symmadd}
\bea
\bG(DD^{*})_{ij} & = & \bG(CC^{*})_{ij} \ ,
\label{symmaddDD}
\\
\bG_{\pm}(CD^{*})_{ij} & = & \bG_{\pm}(DC^{*})_{ij} \ ,
\label{symmaddCD}
\eea
\ees
and the $(2 \times 2)$ $D$-$C$ interference matrices $\bG_{\pm}(CD^{*})$ become
self-adjoint, too.

\section{Partial decay widths of neutrino $N$}
\label{app2}

The formulas for the leptonic decay and semimesonic decay widths
of a sterile Majorana neutrino $N$ have been obtained
in Ref.~\cite{Atre} (Appendix C there), for the masses $M_N \alt 1$ GeV.
Nonetheless, for the higher values of the masses $M_N$, the 
calculation of the semihadronic decay widths becomes increasingly complicated
because not all the resonances are known. Therefore, in Refs.~\cite{HKS,GKS}
an inclusive approach was proposed for the calculation of the total
contribution of the semihadronic decay width of $N$, by replacing the
various (pseudoscalar and vector) meson channels by quark-antiquark channels.
This inclusive approach, based on duality, was applied for high
masses $M_N \geq M_{\eta^{'}} \approx 0.958$ GeV. Here we summarize
the formulas given in Ref.~\cite{HKS} for the decay width channels
(see also: \cite{Atre}).
The leptonic channels are:
\bes
\label{GNlept}
\bea
2 \Gamma(N \to \ell^- \ell^{'+} \nu_{\ell^{'}}) & = & 
|B_{\ell N}|^2 \frac{G_F^2}{96 \pi^3} M_N^5 I_1(y_{\ell},0, y_{\ell^{'}})
(1 - \delta_{\ell \ell^{'}} ) \ ,
\label{GNlept1}
\\
\Gamma(N \to \nu_{\ell} \ell^{'-} \ell^{'+}) & = &
|B_{\ell N}|^2 \frac{G_F^2}{96 \pi^3} M_N^5 
{\big [}
(g_L^{(\rm lept)} g_R^{(\rm lept)}
+ \delta_{\ell \ell^{'}} g_R^{(\rm lept)}) I_2(0,y_{\ell^{'}},y_{\ell^{'}})
\nonumber\\
&&
+ \left( (g_L^{(\rm lept)})^2 + (g_R^{(\rm lept)})^2 
+  \delta_{\ell \ell^{'}}
(1 + 2 g_L^{(\rm lept)}) \right) I_1(0,y_{\ell^{'}},y_{\ell^{'}}) 
{\big ]} \ ,
\label{GNlept2}
\\
\sum_{\nu_{\ell}} \sum_{\nu^{'}} \Gamma(N \to \nu_{\ell} \nu^{'} {\bar \nu}^{'})
& = &
\sum_{{\ell}} |B_{\ell N}|^2 \frac{G_F^2}{96 \pi^3} M_N^5 \ .
\label{GNlept3}
\eea
\ees
In Eq.~(\ref{GNlept1}) factor 2 was included because both decays
$N \to \ell^- \ell^{'+} \nu_{\ell^{'}}$ and $N \to \ell^+ \ell^{'-} \nu_{\ell^{'}}$
contribute ($\ell \not= \ell^{'}$).

If $M_N < M_{\eta^{'}} \approx 0.968$ GeV, the following semimesonic
decays contribute, involving presudoscalar ($P$) and vector ($V$)
mesons: 
\bes
\label{GNmes}
\bea
2 \Gamma(N \to  \ell^- P^+) & = & 
|B_{\ell N}|^2 \frac{G_F^2}{8 \pi} M_N^3 f_P^2 |V_P|^2 F_P(y_{\ell}, y_P) \ ,
\label{GNMesPch}
\\
\Gamma(N \to  \nu_{\ell} P^0) & = & 
|B_{\ell N}|^2 \frac{G_F^2}{64 \pi} M_N^3 f_P^2 (1 - y_P^2)^2 \ ,
\label{GNMesP0}
\\
2 \Gamma(N \to  \ell^- V^+) & = & 
|B_{\ell N}|^2 \frac{G_F^2}{8 \pi} M_N^3 f_V^2 |V_V|^2 F_V(y_{\ell}, y_V) \ ,
\label{GNMesVch}
\\
\Gamma(N \to  \nu_{\ell} V^0) & = & 
|B_{\ell N}|^2 \frac{G_F^2}{2 \pi} M_N^3 f_V^2 \kappa_V^2 (1 - y_V^2)^2 
(1 + 2 y_V^2) \ ,
\label{GNMesV0}
\eea
\ees
where factor $2$ in the charged meson channels is taken because
both decays $N \to  \ell^- M^{'+}$ and $N \to  \ell^+ M^{'-}$ contribute
($M^{'}=P, V$). The factors $V_P$ and $V_V$ are the corresponding CKM
matrix elements involving the valence quarks of the mesons;
and $f_P$ and $f_V$ are the corresponding decay constants. The pseudoscalar
mesons which may contribute are: 
$P^{\pm} = \pi^{\pm}, K^{\pm}$; $P^0 = \pi^0, K^0, {\bar K}^0, \eta$. 
The vector mesons which may contribute are: 
$V^{\pm} = \rho^{\pm}, K^{* \pm}$; $V^0= \rho^0, \omega, K^{*0}, {\bar K}^{*0}$.\footnote{
For the values of the decay constants $f_P$ and $f_V$, see, e.g., Table 1 in
Ref.~\cite{HKS}.}
When $M_N \geq M_{\eta^{'}}$ ($=0.9578$ GeV), the above semimesonic
decay modes are replaced \cite{HKS}, in the spirit of duality, with the
following quark-antiquark decay modes:
\bes
\label{GNquark}
\bea
2 \Gamma(N \to  \ell^- U {\bar D}) & = & 
|B_{\ell N}|^2 \frac{G_F^2}{32 \pi^3} M_N^5 |V_{UD}|^2 I_1(y_{\ell},y_U,y_D) \ ,
\label{GNquark1}
\\
\Gamma(N \to  \nu_{\ell} q {\bar q}) & = & 
|B_{\ell N}|^2 \frac{G_F^2}{32 \pi^3} M_N^5 
\left[
g_L^{(q)} g_R^{(q)} I_2(0,y_q,y_q) +
\left( (g_L^{(q)})^2 + (g_R^{(q)})^2 \right) I_1(0,y_q,y_q)
\right] \ .
\label{GNquark2}
\eea
\ees
In the formulas (\ref{GNlept})-(\ref{GNquark}) we denoted
$y_{x} \equiv M_X/M_N$ ($X = \ell, \nu_{\ell},  P, V, q$), and
in Eqs.~(\ref{GNquark}) we denoted: $U=u,c$; $D=d,s,b$; $q=u,d,c,s,b$.
The values of quark masses which we used were: $M_u=M_d = 3.5$ MeV;
$M_s=105$ MeV; $M_c=1.27$ GeV; $M_b=4.2$ GeV.
The SM neutral current couplings in Eqs.~(\ref{GNlept2}) and (\ref{GNquark2})
are
\bes
\label{NCc}
\bea
g_L^{(\rm lept)} &=& - \frac{1}{2} + \sin^2 \theta_W \ ,
\quad 
g_R^{(\rm lept)} =\sin^2 \theta_W \ ,
\label{NCcl}
\\
g_L^{(U)} & = & \frac{1}{2} - \frac{2}{3} \sin^2 \theta_W \ ,
\quad
g_R^{(U)} = - \frac{2}{3} \sin^2 \theta_W \ ,
\label{NCcU}
\\
g_L^{(D)} & = & - \frac{1}{2} + \frac{1}{3} \sin^2 \theta_W \ ,
\quad
g_R^{(U)} =  \frac{1}{3} \sin^2 \theta_W \ .
\label{NCcD}
\eea
\ees
The neutral current couplings $\kappa_V$ of the neutral vector mesons are
\bes
\label{kV}
\bea
\kappa_V & = & \frac{1}{3} \sin^2 \theta_W  \quad (V=\rho^0, \omega) \ ,
\label{kV1}
\\
\kappa_V & = & - \frac{1}{4} +\frac{1}{3} \sin^2 \theta_W  
\quad (V=K^{*0}, {\bar K}^{*0}) \ .
\label{kV2}
\eea
\ees
The kinematical expressions $I_1$, $I_2$, $F_P$ and $F_V$ are
\bes
\label{kinex}
\bea
I_1(x,y,z) & = & 12 \int_{(x+y)^2}^{(1-z)^2} \; \frac{ds}{s} 
(s - x^2 - y^2) (1 + z^2 -s) 
\lambda^{1/2}(s,x^2,y^2) \lambda^{1/2}(1,s,z^2) \ ,
\label{I1}
\\
I_2(x,y,z) & = & 24 y z \int_{(y+z)^2}^{(1-x)^2} \; \frac{ds}{s} 
(1 + x^2 - s) 
\lambda^{1/2}(s,y^2,z^2) \lambda^{1/2}(1,s,x^2) \ ,
\label{I2}
\\
F_P(x,y) & = & \lambda^{1/2}(1,x^2,y^2) \left[(1 + x^2)(1 + x^2-y^2) - 4 x^2
\right] \ ,
\label{FP}
\\
F_V(x,y) & = & \lambda^{1/2}(1,x^2,y^2) \left[(1 - x^2)^2 + (1 + x^2) y^2 - 2 y^4 \right] \ ,
\label{FV}
\eea
\ees
where $\lambda$ function is written in Eq.~(\ref{lambda}).
Using these formulas, the total decay width $\Gamma(N_j \to {\rm all})$
can be calculated, and coefficients ${\cal N}_{\ell N_j}$
of Eq.~(\ref{calK}) at the mixing terms $|B_{\ell N_j}|^2$ can be evaluated 
and are presented in Fig.~\ref{FigcNellN}. The small kink in the curves
of Fig.~\ref{FigcNellN} at $M_N=M_{\eta^{'}}$ ($=0.9578$ GeV) appears due to 
the replacement there (i.e., for $M_N \geq M_{\eta^{'}}$) of the
semihadronic decay channel contributions
by the quark-antiquark channel contributions; we see that the
duality works quite well there, with the exception of the case $\ell = \tau$
because of the large $\tau$ lepton mass.

\section{Explicit expression for the function $Q$}
\label{app3}

The expression (\ref{GDD}) can be obtained by using in the integration
over the phase space of three final particles [Eqs.~(\ref{GM1})-(\ref{d3})],
for the contribution of the $N_j$ neutrino, the identity
\bes
\label{d3d2}
\bea
\lefteqn{
d_3 \left( M(p_M) \to  \ell_1(p_1) \ell_2(p_2) M^{'}(p_{M'}) \right) 
}
\nonumber\\
&=& d_2 \left( M(p_M) \to  \ell_1(p_1) N_j(p_N) \right) d p_N^2
d_2 \left( N_j(p_N) \to  \ell_2(p_2) M^{'}(p_{M'}) \right)
\label{d3d2D}
\\
&=& d_2 \left( M(p_M) \to  \ell_2(p_2) N_j(p_N) \right) d p_N^2
d_2 \left( N_j(p_N) \to  \ell_1(p_1) M^{'}(p_{M'}) \right) \ ,
\label{d3d2C}
\eea
\ees
where the first identity can be used for the $DD^{*}$ contribution
(where $p_N=p_M-p_1$) and the second for the $CC^{*}$ contribution
(where $p_N=p_M-p_2$). Using the identity (\ref{P1P1}) in the
$DD^{*}$ contribution, and the analogous identity for the $CC^{*}$
contribution, the integration over $d p_N^2$ becomes trivial,
and the $d_2$-type of integrations are straightforward.\footnote{
This is equivalent to the factorization approach $\Gamma(M \to \ell_1 N_j)
{\rm Br}(N_j \to \ell_2 M^{'})$ valid when $N_j$ is on-shell.}
The resulting expression for $\bG(DD^{*})_{jj}$ is then the
expression Eq.~(\ref{GDD}) with the notations (\ref{notGDD}) and (\ref{notx}),
where the function $Q$ has the form
\bea
Q(x; x_{\ell_1}, x_{\ell_2}, x') & = & 
{\bigg \{} \frac{1}{2}(x - x_{\ell_1})(x - x_{\ell_2})(1 - x - x_{\ell_1})
\left( 1 - \frac{x'}{x} +\frac{x_{\ell_2}}{x} \right) 
\nonumber\\
&& +  {\big [} 
- x_{\ell_1} x_{\ell_2} (1 + x' + 2 x - x_{\ell_1} - x_{\ell_2} )
- x_{\ell_1}^2 (x - x') + x_{\ell_2}^2 (1 - x) 
\nonumber\\
&&
+ x_{\ell_1} (1+x) (x - x') - x_{\ell_2} (1-x)(x+x') {\big ]}
{\bigg \}} \ .
\label{Q}
\eea

\end{document}